\documentclass{aa}  

\bibpunct{(}{)}{;}{a}{}{,} 

\usepackage{graphicx}
\usepackage{lscape}

\usepackage{txfonts}
\usepackage{hyperref}
\begin{document}

   \title{
Detection of new O-type stars in the obscured stellar cluster Tr\,16-SE in the Carina Nebula
with KMOS \thanks{Based on observations collected at the European Southern Observatory under 
ESO program 097.C-0102.}}

   \author{T.~Preibisch\inst{1}
          \and
          S.~Flaischlen\inst{1}
          \and
          C.~G\"oppl\inst{1}
          \and
          B.~Ercolano\inst{1}
          \and
          V.~Roccatagliata\inst{2,3,4}
          }

   \institute{Universit\"ats-Sternwarte M\"unchen, 
              Ludwig-Maximilians-Universit\"at,
              Scheinerstr.~1, 81679 M\"unchen, Germany\\
              \email{preibisch@usm.uni-muenchen.de} 
         \and
             Dipartimento di Fisica ``E. Fermi'', Universita di Pisa
             Largo Bruno Pontecorvo 3, 56127 Pisa, Italy
         \and
          INAF-Osservatorio Astrofisico di Arcetri, Largo E.~Fermi 5, 
           50125 Firenze, Italy
         \and
          INFN, Sezione di Pisa, Largo Bruno Pontecorvo 3, 56127 Pisa, Italy
             }

\titlerunning{KMOS detection of new O-type stars in the obscured cluster Tr\,16-SE}
\authorrunning{Preibisch et al.}

   \date{Received 7 October 2020; accepted 11 February 2021}


  \abstract
{ 
The Carina Nebula harbors a large population of high-mass stars,
including at least 75 O-type and Wolf-Rayet (WR) stars, 
but the current census is not complete since further
high-mass stars may be hidden in or behind the dense dark clouds that pervade the association.}
{With the aim of identifying optically obscured O- and early B-type stars in the 
Carina Nebula, we performed the
first infrared spectroscopic study of stars in the optically obscured stellar
cluster Tr\,16-SE, located behind a dark dust lane south of $\eta$ Car.
}
{We used the integral-field spectrograph KMOS at the ESO VLT to obtain 
$H$- and $K$-band spectra
with a resolution of $R \approx 4000$ ($\Delta \lambda \approx 5 \AA$)
for 45 out of the 47 possible OB candidate stars in Tr\,16-SE, and we
derived spectral types for these stars.}
{We find 15 stars in Tr~16-SE with spectral types between O5 and B2
(i.e.,~high-mass stars with $M \ge 8\,M_\odot$), only two of which
were known before.
An additional nine stars are classified as (Ae)Be stars (i.e.,~intermediate-mass
pre-main-sequence stars), and 
most of the remaining targets show clear signatures of being late-type stars
and are thus most likely foreground stars or background giants
unrelated to the Carina Nebula. Our estimates of the stellar luminosities
 suggest that nine of the 
15 O- and early B-type stars are members of Tr\,16-SE, whereas the other six seem
to be background objects.
 }
{Our study increases the number of spectroscopically identified 
high-mass stars ($M \ge 8\,M_\odot$) in Tr~16-SE from two to nine and
shows that Tr~16-SE is one of the larger clusters in the Carina Nebula.
Our
identification of three new stars with spectral types between
O5 and O7 and four new stars with spectral types O9 to B1
significantly increases the number of spectroscopically identified O-type stars
in the Carina Nebula.
}
                {}


   \maketitle
%

\section{Introduction}

The Carina Nebula, at a distance of $\approx 2.3$~kpc
\citep[see][for a review]{SB08},
is one of the most massive and active star forming regions 
in our Galaxy and harbors a large population of high-mass stars ($M \ge 8\,M_\odot$). With 70 optically identified O-type stars, three known Wolf-Rayet (WR)  stars, and
$\eta$ Car with its companion
\citep[see][]{Smith06,2016AJ....152..190A,2018AJ....155..190H},
the currently known population
comprises 75 stars with masses above $18\,M_\odot$.
These massive stars constitute a large OB association, 
in which Tr~14, 15, and 16 are the most prominent open clusters 
in optical images.

\begin{figure*}
\parbox[t]{18.0cm}{
\parbox[c]{8.9cm}{\includegraphics[width=8.9cm]{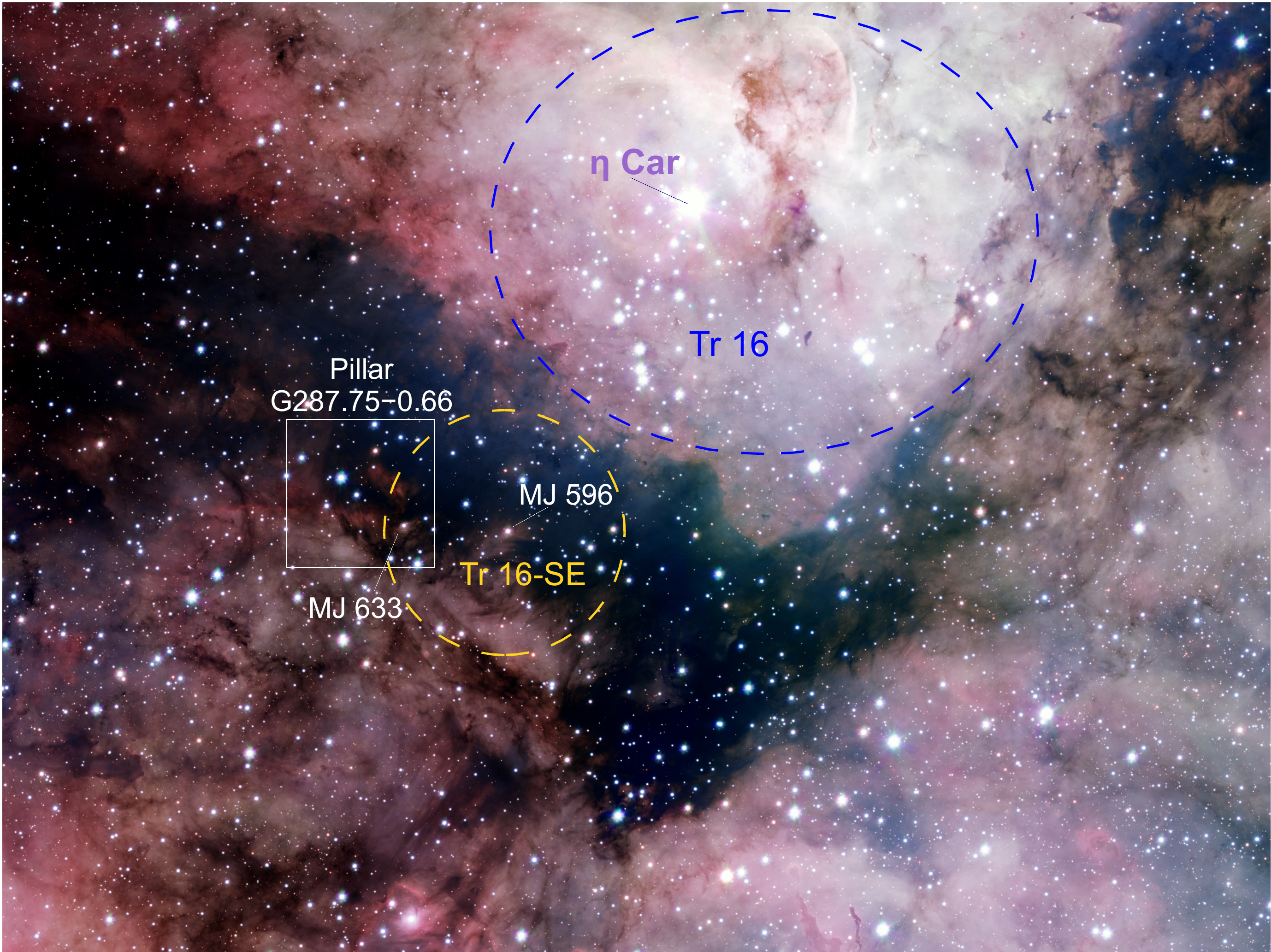}}\hspace{3mm}
\parbox[c]{8.9cm}{\includegraphics[width=8.9cm,trim=75 57 75 57,clip]{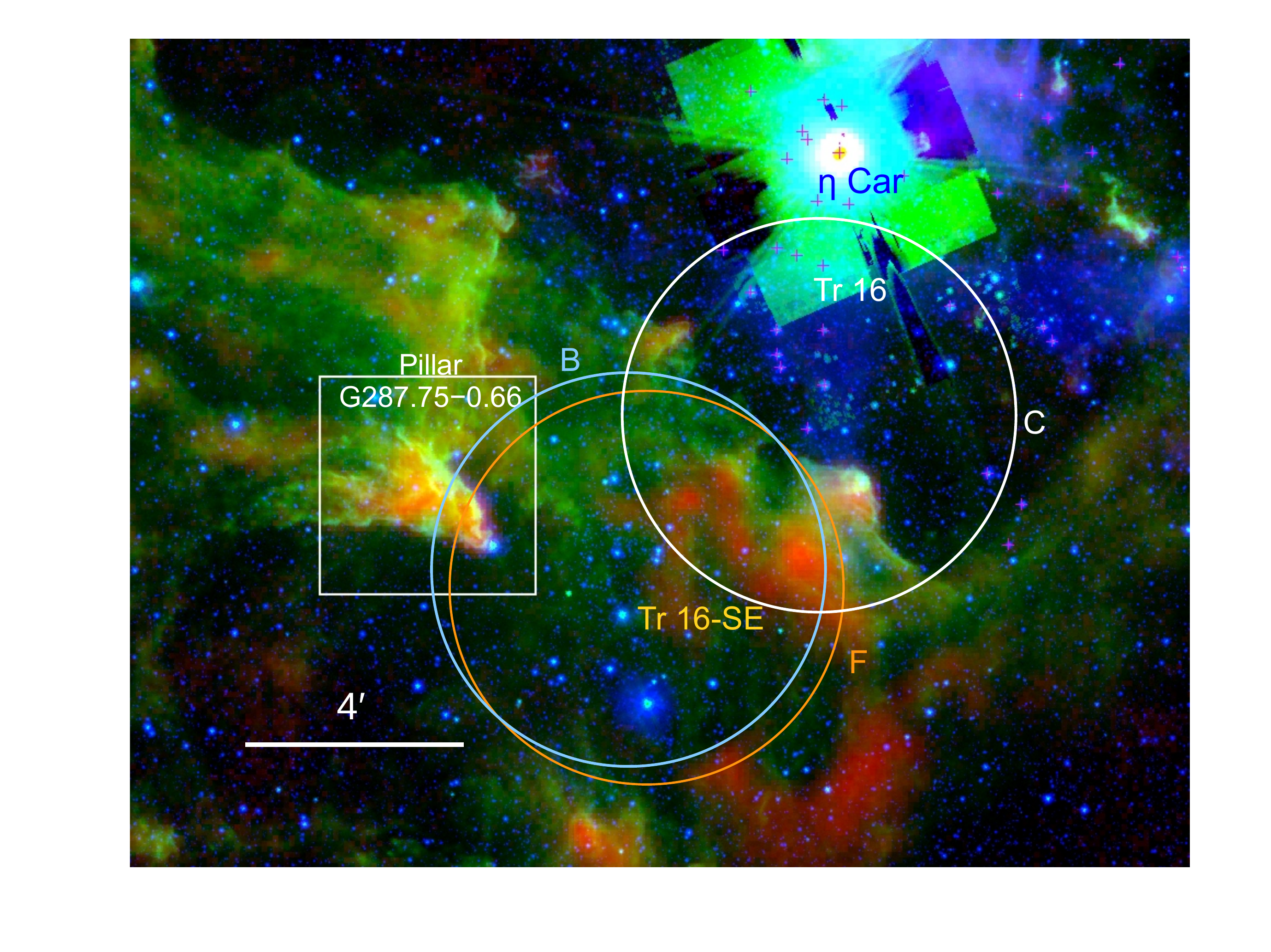}}}
\caption{Optical and infrared images of the central part of the Carina Nebula.
Left:
Optical image reproduced from the ESO Photo Release eso1250
(image credit: VPHAS+ Consortium/Cambridge Astronomical Survey Unit).
It is composed of a $B$-band image in the blue color channel, a $g$-band image in the green color channel,
and $r$-band and H$\alpha$-band images in the red color channel.
The white box marks the region around the
pillar G~287.75$-0.66$ (which is not visible in this optical image).
The dashed yellow circle marks the region of the obscured cluster Tr\,16-SE
and the dashed blue ellipse the cluster Tr~16. The positions of the
stars MJ~596, MJ~633, and  $\eta$~Car are marked. North is up, and east is to the left. Right: Color composite image of the region around the pillar,
composed from our VISTA $K_s$-band image in blue,
the \textit{Spitzer} $5.8\,\mu$m image in green
(where the area around $\eta$~Car [near the upper right corner]
is affected by saturation),
and our LABOCA $870\,\mu$m map in red.
The white $4' \times 4'$ square marks the region of the pillar G~287.75$-0.66$, and
the circles show the KMOS patrol fields for the \textbf{B} (blue), \textbf{F} (orange),
and \textbf{C} (white) observations.
North is up, and east is to the left.
\label{Opt-IR.fig}}
\end{figure*}

The Carina Nebula still contains a very large mass 
\citep[$\sim 10^6\,M_\odot$;][]{Preibisch12} of dusty gas clouds.
Maps of the cloud column density derived from far-infrared \citep{Preibisch12}, 
submillimeter \citep{CNC-Laboca}, and CO observations
\citep{Rebolledo16}
show that the structure of these clouds is very inhomogeneous and that the
visual cloud extinction reaches values of $A_V \sim 10$~mag or more 
at several locations in the nebula.
Given the considerable obscuration effect of these dark clouds,
only deep infrared observations can reveal the full stellar population. 
Deep near-infrared (NIR) surveys of the Carina Nebula have been
obtained with HAWK-I on the  Very Large Telescope (VLT) at the European Southern Observatory  \citep[ESO;][]{HAWKI-survey} 
as well as with the VISTA telescope 
\citep{VISTA1}.
The Carina Nebula was also mapped at mid-infrared wavelengths with the
 \textit{Spitzer} Space Observatory \citep{Smith10b}.
In combination with the comprehensive X-ray observations obtained
 in the \textit{Chandra Carina Complex Project} 
\citep[CCCP; see][]{CCCP-intro},
these infrared data yielded comprehensive information
about the stellar populations \citep[e.g.,][]{CCCP-Clusters, CCCP-HAWKI, Povich11a}.

However, the current census of high-mass stars
in the Carina Nebula is probably still substantially incomplete.
For example, \citet{Roccatagliata13}
found that the total far-infrared luminosity of the cloud complex
is $\approx 70\%$ larger than expected from the total stellar far-ultraviolet luminosity of the
currently known OB stars.
The analysis of bright \textit{Spitzer} infrared sources
by
\citet{Povich11a} led to the identification of 94 new candidate OB stars 
in the Carina Nebula with extinction values up to $A_V \approx 35$~mag.
 
Due to the location of the Carina Nebula very close to the galactic plane,
the vast majority of all detected infrared sources are distant field
stars in the galactic background.
The only way to reliably identify obscured OB stars among the many background contaminators
is thus by spectroscopy.
Spectroscopic surveys \citep[see the summary in][for earlier observations]{Smith06}
have, however, so far only been performed for the
 unobscured population of optically bright stars \citep{2016AJ....152..190A,2017A&A...603A..81D,2017MNRAS.465.1807M,10.1093/mnras/sty748}.
This lack of spectroscopic information for the obscured, optically faint 
high-mass star candidates is a serious gap in our knowledge of the
stellar population in the Carina Nebula and limits our understanding of the
global energetics and feedback processes in this important star forming region. 
\medskip

The {\sf V}-shaped dark cloud structure that intersects the 
central parts of the Carina Nebula contains a clustering of 
infrared and X-ray sources, denoted as Tr\,16-SE
\citep{2007ApJ...656..462S,Smith10b,CCCP-Clusters}, about 9 arcminutes southeast  of $\eta$~Car and the optically prominent
stellar cluster Tr~16 (see Fig. \ref{Opt-IR.fig}). 
Most stars in Tr~16-SE are invisible or very faint in optical images,
and until recently the only star with a known optical spectral type
was 
the  double-lined O5.5V+O9.5V spectroscopic binary MJ~596 \citep{Niemela06},
also known as V662~Car.
The study of the bright \textit{Spitzer} point sources 
by \citet{Povich11a} found six candidate OB stars in Tr~16-SE,
but none of these have been confirmed with spectroscopy to date. This makes Tr~16-SE a very interesting target
for the search for still unidentified high-mass stars in  the Carina Nebula.
\medskip

The very high levels of ionizing radiation and stellar wind power
from  the numerous massive stars in the Carina Nebula
profoundly influence the surrounding clouds, making it a
textbook example for studies of stellar feedback and the resulting processes
of cloud disruption and triggered star formation.
The infrared to submillimeter images 
\citep[see][]{Smith10b,Preibisch12,CNC-Laboca}
revealed numerous pillar-like cloud structures in the southern parts of the Carina Nebula.
Such pillars are thought to be a natural outcome 
of the feedback from strong ionizing radiation fields on clouds
\citep[see, e.g.,][]{Gritschneder10,2013MNRAS.430..234D,McLeod16,Klaassen20}.
While almost all pillars in the Carina Nebula point toward the massive stars around $\eta$~Car in Tr~16,
the \textit{Spitzer} image (see Fig.~\ref{Opt-IR.fig}) reveals 
one prominent exception approximately $9\arcmin$ southeast of $\eta$~Car,
where one bright and well-structured  pillar does not point 
toward $\eta$~Car (which would be north)
but instead points in a southwestern direction  toward a region where no particularly bright stars are seen in
optical images. This pillar was denoted as G~287.75$-0.66$ by \citet{Smith10b} and 
is part of the cloud structure that 
creates the remarkable \textsf{V}-shaped dark lane in the
central part of the Carina Nebula in optical images (see Fig.~1).
Interestingly, this pillar only points in the direction of
the obscured cluster Tr~16-SE.

\medskip

Our aims are thus 
to perform the first systematic spectroscopic study of the
stellar population in Tr~16-SE and thus improve the census of massive stars ($M \ge 8\,M_\odot$) 
in the Carina Nebula;
we are therefore interested in identifying stars with spectral types from
 O to B2, which we call ``OB2 stars'' in the following.
Due to the high level of obscuration, most stars in Tr~16-SE
are too faint for optical spectroscopy; we therefore 
obtained NIR spectra for a magnitude-limited sample of
50 stars in the cluster. 

\section{K-band Multi-Object Spectrograph observations \label{observations.sec}}

\subsection{Target sample and observing strategy \label{strategy.sec}}

Our goal was to obtain $H$- and $K$-band
infrared spectra of all potential high-mass stars in Tr~16-SE ($M \ga 8\,M_\odot$, corresponding to spectral types
from O to B2) in the Carina Nebula.
The target stars were selected from our VISTA photometry catalog
of the Carina Nebula \citep[see][]{VISTA1}.
A brightness limit of $K_s \leq 12$~mag was chosen because
any star at the distance of the Carina Nebula with a mass of 
$M \ge 8\,M_\odot$ [$M \ge 18\,M_\odot$]
will be brighter than this limit for extinctions of up to $A_V \approx 11$~mag
[$A_V \approx 28$~mag].
We can thus be confident that such a magnitude-limited sample will 
contain all the O-type stars ($M \ge 18\,M_\odot$) and most of the early B-type stars 
($8\,M_\odot \le M \le 18\,M_\odot$) in Tr~16-SE. We did not use any further conditions (e.g., luminosity estimates
from spectral energy distribution analysis or X-ray detection)
in the construction of our sample
in order to keep it as unbiased as possible.

As the spatial selection region for our Tr~16-SE sample, we used
a $5.7\arcmin \times 5\arcmin$ box
centered at the J2000 coordinates $\mathrm{R.A.} = 10^{\rm h}\,45^{\rm m}\,31.7^{\rm s}$ and
$\mathrm{Dec} = -59\degr\,48\arcmin\,48\arcsec$.
This area contains 47 stars\footnote{We excluded the
star BM~VII~10 (= 2MASS J10453185-5951094)
because it is known to be an S~star \citep[see][]{2000A&AS..145...51V},
i.e.,~a cool giant,
and would have been much too bright $(K_s = 5.0$) for our observations.
}
 with $K_s \le 12$.
We were able to obtain spectra of 45 of these 47 stars (i.e., we~have an
observational completeness of 95.74\%);
the only two stars in our selection region that could
not be observed are
VCNS J104554.89-594818.8 ($K_s = 7.84$) 
and
VCNS J104554.36-595006.6 ($K_s = 11.75$).

Due to the number of our targets and the wide range of magnitudes
of the individual target stars, the
observational sequence was split into three individual observations,
denoted as the \textbf{B} (bright star), \textbf{F} (faint star), and \textbf{C} (calibrator star) observations in the following.
The \textbf{B} observation contained the stars in Tr~16-SE
with $K_s \approx [7.0 - 10.6]$~mag and $H \approx [7.5 - 11.7]$~mag.
The \textbf{F} observation contained the stars in Tr~16-SE
with $K_s \approx [10.6 - 12.0]$~mag and $H \approx [10.8 - 13.2]$~mag.

\begin{table*}
\caption{Calibrator star sample.}             
\label{calibrator-tab}      
\centering                          
\begin{tabular}{l c l l l}        
\hline\hline                 
KMOS & J2000 coordinates &  Star & \multicolumn{2}{c}{Spectral type}\\    
spectrum & [hh:mm:ss.ss$-$dd:mm:ss.s] &  Name & [Smith] & [Sota]\\    
\hline                        
C2  & 10:44:41.76$-$59:46:56.3 &  CPD-59 2600  & O6 V((f)) & O6 V((f)) \\
C3  & 10:44:47.29$-$59:43:53.2 &  CPD-59 2603A & O7 V((f)) & O7.5 V \\
C4  & 10:45:12.71$-$59:44:46.0 &  CPD-59 2635A & O8 V    & O8 V\\
C5  & 10:45:12.87$-$59:44:19.2 &  CPD-59 2636A & O7 V    & O8 V \\
C6  & 10:45:16.51$-$59:43:37.0 &  CPD-59 2641  & O5 V    & O6 V \\
C7  & 10:45:12.21$-$59:45:00.3 &  HD 93343     & O7 V(n) &  O8 V \\
C8  & 10:45:05.84$-$59:43:07.7 &  Tr~16-9      & O9.5 V  & O9.7 IV \\
C9  & 10:45:08.22$-$59:46:06.9 &  Tr~16-22     & O8.5 V  & O8.5 V\\
C10 & 10:45:05.79$-$59:45:19.5 &  Tr~16-23     & O7 V    & O7.5 V \\
C11 & 10:45:05.87$-$59:44:18.8 &  Tr~16-24     & B2 V    & --\\
C12 & 10:45:09.74$-$59:42:57.2 &  Tr~16-74     & B1 V    & --\\
C13 & 10:45:12.65$-$59:42:48.7 &  Tr~16-76     & B2 V    & --\\
C14 & 10:45:20.57$-$59:42:51.2 &  Tr~16-115    & O8.5 V  & O9 V \\
C15 & 10:44:59.90$-$59:43:14.9 &  CPD-59 2618  & \multicolumn{2}{c}{B1.5 V}\\
\hline                                   
\end{tabular}
\tablefoot{The spectral types listed in Col. 4 are from \citet{Smith06}, and
those in Col. 5 are from \citet{Sota_2014}; the spectral type of \mbox{CPD-59 2618}
 is from \citet{MJ93}.}
\end{table*}

The  \textbf{C} observation targeted
14 known high-mass stars in the cluster Tr~16, for which
spectral types between O5 and B2 had been previously determined from
optical spectroscopy; Table \ref{calibrator-tab} lists the positions, names, and
spectral types of these stars.
They were observed and analyzed in the same way as our target stars
in Tr~16-SE and then used as spectral standards for the classification
of the target stars in Tr~16-SE.
Figure \ref{Opt-IR.fig} shows the positions of the three observations.

\subsection{KMOS configuration}

The \textit{K-band Multi-Object Spectrograph} (KMOS) at the ESO VLT
allows the simultaneous observation of up to 24 target stars within
a circular $7.2\arcmin$ diameter patrol field.
As described in more detail in \citet{2013Msngr.151...21S} and
on the ESO instrument web page\footnote{https://www.eso.org/sci/facilities/paranal/instruments/kmos.html},
KMOS employs 24 configurable arms that position pickoff mirrors at
user-specified locations in the focal plane.
Each subfield is fed to an image slicer integral-field unit (IFU)
with a square field of view of $2.8\arcsec \times 2.8\arcsec$ that
provides a uniform spatial sampling of $0.2\arcsec \times 0.2\arcsec$.
The light from each IFU
is then dispersed by grating spectrometers that generate
$14 \times 14$ spectra with $\sim 1000$ Nyquist-sampled spectral resolution
elements.

While KMOS allows a rather flexible spatial placement of the
24 individual pickoff mirrors, there are some restrictions on the configuration, for example,~a minimum distance of at least $6\arcsec$ 
between two neighboring pickoff mirrors.
It is thus not always possible to observe all desired target stars in one field.
After performing detailed experiments, we were finally able to establish 
configurations for the three observations that allowed us to
obtain spectra of 45 out of the 47 OB candidate stars in Tr~16-SE 
and all 14 calibrator stars.

The resulting spectra are denoted by the corresponding \textbf{B}, \textbf{F}, or
\textbf{C} prefixes and a running number corresponding to the IFU (e.g., B1).
While the target stars were generally centered on the IFU,
our target star, J104530.29--594824.5, has a close companion
(J104530.42--594825.4, angular separation $1.3\arcsec$), and as such 
both stars could be observed in one
IFU  (F12); their spectra are denoted as F12-1 and F12-2.
In order to employ all available KMOS arms in the \textbf{F} observation,
we also observed one star (F21)
with $K_s = 12.24$, which is slightly fainter than our magnitude limit.

Since the \textbf{C} observation partly 
overlapped with the Tr~16-SE target field (see Fig.~\ref{Opt-IR.fig}), 
six of the KMOS arms not needed for the calibrator stars 
were used for additional, redundant observations of target stars in Tr~16-SE;
the two spectra of these double observations
were reduced and analyzed
independently, and they served as a consistency check for our analysis.
Furthermore, three KMOS arms (C16, C17, and C18) were put on stars in or near 
the cluster Tr~16.

\begin{table*}
\caption{Parameters of the KMOS observations.}             
\label{aimpoints-tab}      
\centering                          
\begin{tabular}{l l l l }        
\hline\hline                 
Observation & Aimpoint (J2000)& Obtained Spectra & Exposure Time\\    
\hline                        
\textbf{B} & 10:45:35.0 $-$59:48:45 & 23 target stars & $ 14 \times 3\,{\rm s} = 42$~s (H), $13 \times 3\,{\rm s} = 39$~s (K) \\
\textbf{F} & 10:45:31.7 $-$59:49:07 & 24 target stars & $ 8 \times 20\,{\rm s} = 160$~s (H), $ 6 \times 20\,{\rm s} = 120$~s (K) \\
\textbf{C} & 10:45:07.6 $-$59:45:52 & 14 spectral calibrators + 9 target stars & $ 14 \times 3\,{\rm s} = 42$~s (H), $13 \times 3\,{\rm s} = 39$~s (K) \\
\hline                                   
\end{tabular}
\end{table*}

In each observation, one KMOS arm was used
for a simultaneous sky observation.
The coordinates of the stars in our three different observations (\textbf{B}, \textbf{F}, and \textbf{C})
 are listed in Table~\ref{results.table}.

\subsection{Spectroscopic observations}

We obtained KMOS spectra in the $H$  and the $K$ spectral bands, which provide
a wavelength coverage of $[1.456 - 1.846]\,\mu$m and
$[1.934 - 2.460]\,\mu$m, respectively, and a spectral resolving power of
$R(H) \simeq 4040$ and $R(K) \simeq 4230$.
The observations were carried out on 21 February 2016 (PI: Preibisch, 097.C-0102(A))
between 02:37:34 UT and 04:58:25 UT 
under photometric sky conditions.
The median Differential Image Motion Monitor (DIMM) seeing  
at $500~\mathrm{nm}$ was $\approx 1.4\arcsec$; the
corresponding full width at half maximum (FWHM) of the point spread function (PSF) in the NIR is $\approx 1.0\arcsec$. 
For each observation,
target and sky-offset exposures were alternated in a target-sky-target sequence. 
The sky regions were observed  by rotating and shifting 
the arms without altering the configuration.
Each science observation was performed with three dither positions and a dither size 
of $0.2\arcsec$. 

For the \textbf{B} and \textbf{C} observations, 
the individual integration time was $3$ seconds
and the number of integrations was $14$ ($13$) for the $H$-band ($K$-band), 
resulting in a total exposure time of $42~\mathrm{s}$  ($39~\mathrm{s}$). 
Furthermore, an additional long exposure on sky of $42$ ($39$) seconds 
for the $H$-band ($K$-band) was executed at the end of each science observation.
For the \textbf{F} observation,  the individual integration time was $20$ seconds; the 
number of integrations was eight (six) for the $H$-band ($K$-band), 
leading to a total exposure time of $160~\mathrm{s}$ ($120~\mathrm{s}$).

\subsection{Data reduction}

The data were reduced using the ESO Recipe Flexible Execution Workbench (Reflex) and 
the ESO Recipe Execution Tool (esorex),
as described by \citep{Freudling13}, using the standard KMOS/esorex routines.
Sky subtraction
and telluric correction were performed with the sky tweaking method
described by \citet{2013A&A...558A..56D};
a comprehensive description of the method can be found in \citet{2019Msngr.177...14C}. 
The standard star used for the telluric correction (and later also for flux calibration)
was the B3V star HD~93695.

From the final data cubes, spectra were extracted using the ``optimal extraction method'' provided
by the pipeline. For the two cases where more than one source was present per IFU
(targets F12-1 \& F12-2 and targets B24 \& F24),
the extraction was performed manually with QFitsViewer, employing
circular apertures with radii of $2~\mathrm{pixels}$ and $3~\mathrm{pixel}$s.

\subsection{Resulting spectra}

The resulting spectra were flux-calibrated using the B3V standard star
and stored in units of $\mathrm{erg}\, \mathrm{s}^{-1}\, \mathrm{cm}^{-2}\, \AA^{-1}$.
They have a spectral dispersion of  
$2.16\,\AA$ per pixel in the $H$-band and
$2.81\,\AA$ per pixel in the $K$-band.
For the analysis of the line strengths, we performed a continuum normalization 
 of all spectra with cubic spline interpolation
using ten to 30 points in the continuum.
The normalized spectra are displayed in Figs.~\ref{fig:spectra-B-H} to ~\ref{fig:spectra-F-K}.

In order to characterize the quality of our spectra, we determined
the signal-to-noise ratio (S/N) with the DER\_SNR algorithm described in \citet{2008ASPC..394..505S}.
In order to avoid spectral regions with strong features, we used
 the wavelength intervals $[1.62 - 1.63] + [1.65 - 1.67]\,\mu$m for the $H$-band
and  $[2.12 - 2.15] + [ 2.22 - 2.25]\,\mu$m for the $K$-band.
The S/N of the target spectra ranges from values of 14 to 545 with a mean (median)
value of 132 (81) for the $H$-band,
and from values of 26 to 348 with a mean (median) value of 115 (95) for the $K$-band.


\section{Spectral classification\label{spectral-analysis}}

\subsection{Classification strategy}

Since the aim of our study was to identify high-mass stars ($M \ge 8\,M_\odot$) in the Carina Nebula,
we focused our efforts on those  targets that show evidence
of being of O- or early B-type stars (not later than B2),
which we denote as OB2 stars.
The primary spectral indicators for these hot stars 
are the helium lines. It is well established
\citep[see, e.g.,][]{Gray,Hanson05} that stars with
spectral types earlier than B3 generally show 
\ion{He}{I}\, lines and that stars with
spectral types O8 or earlier
also show \ion{He}{II} lines.
These expectations are very well confirmed in the spectra
of our calibration stars that have optically
determined spectral types. 
We therefore considered all stars that clearly show He lines
as OB2 stars and estimated their spectral types as described in Sect. 
\ref{spectral-typing.sec}.

Stars that show no clear He lines must have
spectral types later than B2 and thus stellar masses $< 8\,M_\odot$.
As these stars are not relevant for the aims of our study and because
our target sample is clearly incomplete in this mass range,
we limited our efforts to characterizing such stars with crude estimates.
The $K_s \leq 12$~mag limit for our target sample implies the lower limit 
in stellar mass
(or spectral type) that we can expect to find in our sample if the target star
is at the distance of the Carina Nebula. According to the
CMD 3.3 web interface\footnote{http://stev.oapd.inaf.it/cgi-bin/cmd}
for the PARSEC stellar models \citep{2012MNRAS.427..127B},
this magnitude limit corresponds to a stellar mass of 
$\sim 2.5\,M_\odot$ and a late F to early G spectral type
for stars at ages between 1 and 3 Myr if there is no extinction.
We thus expect that our sample should contain some
intermediate-mass young stars, some of which should display the
typical spectral signature of Herbig AeBe stars \citep[see][]{1994ASPC...62....3H}.

These considerations also imply that any cool stars (spectral type later than G) 
we find in our sample must either be
foreground stars or red giants; in both cases, these stars would be
unrelated ``field stars'' (since no red giants are expected to be present in the
young stellar population of the Carina Nebula).
Therefore, stars with cool spectra are not interesting\footnote{The possibility
of RSG stars (i.e., evolved massive stars) will be addressed in Sect.~\ref{RSG.sec}.}
for the purposes of our study.

\subsection{Diagnostic lines in the KMOS spectra}

For the classification of our target stars we
used the following diagnostic lines in the $H$- and $K$-bands.\smallskip

\subsubsection{{\ion{He}{II} lines}} Our KMOS spectra contain the  
\ion{He}{II}\, 7--12 line at  $1.6918\,\mu$m and the 
\ion{He}{II}\, 7--10 line at  $2.1885\,\mu$m, both of which
are clear signatures of O-type stars. 
In our calibrator sample, the \ion{He}{II}\,7--12 line is clearly present 
in all stars down to spectral type O6 but is not seen 
in later type stars; it shows equivalent widths up to
0.35~$\AA$ in the hottest stars.
The \ion{He}{II}\,7--10 line 
is clearly seen in the hottest stars and down to spectral type O7.5, 
but not in later type stars;
it shows equivalent widths up to 1.0~$\AA$ in the hottest stars.

\subsubsection{{\ion{He}{I} lines}} 
For our analysis, we used the
\ion{He}{I} 1s3p~$^3$P$^0$ $\rightarrow$ 1s4d~$^3$D triplet at $1.7002\,\mu$m, the \ion{He}{I} 1s3p~$^3$P$^0$ $\rightarrow$ 1s4s~$^3$S triplet  
at $2.1120\,\mu$m, and the
\ion{He}{I} 1s3p~$^3$P$^0$ $\rightarrow$ 1s4s~$^1$S line and $2.1132\,\mu$m, 
the second and third of which form a blend at $\approx 2.1126\,\mu$m at
the resolution of our KMOS spectra.
In our calibrator sample, these
\ion{He}{I} lines are seen in all O-type stars and down to spectral type
B2 (the latest spectral type in 
our calibrator sample). The strength of \ion{He}{I} $1.7002\,\mu$m
increases  from 0.5~$\AA$ for the O5 stars toward  1.9 $\AA$ in the early B-type stars.
The \ion{He}{I} $2.1126\,\mu$m line shows a similar tendency of line strength versus 
spectral type,
but it appears in emission (with equivalent widths up to $-3 \,\AA$)
in the hottest (O5) stars. It is blue-shifted
in the stars later than O6.
These \ion{He}{I} lines are expected to disappear in stars cooler than
spectral types B3-4 \citep[see, e.g.,][]{2005A&A...440..261R}.

\subsubsection{{H lines}}
Our spectra contain the Br~7 ($2.1655\,\mu$m) to 
Br~15 ($1.5701\,\mu$m) atomic hydrogen  lines. 
Most of these, in particular the Br~7 (= Br$\gamma$) line,
 are seen in all our calibrator spectra.
As expected, their line strength generally decreases with increasing temperature.
The equivalent widths of the
Br$\gamma$ line in our calibrator spectra range up to $6.5\,\AA$;
the line is seen in emission in one star.
In general, strong Br$\gamma$ lines are also seen in substantially cooler
(late B, A, and F) stars and are expected to disappear only at late G spectral types.

\subsubsection{\ion{Na}{I} and \ion{Ca}{I} lines}
The \ion{Na}{I} lines at $2.2056\,\mu$m  and  $2.2084\,\mu$m
and the \ion{Ca}{I} lines at  $2.2608\,\mu$m + $2.0796\,\mu$m + $2.0836\,\mu$m
are clear signatures of cool stars (spectral type G or later).
They are generally absent in the spectra of our OB2 calibrator stars.
We note that one of our calibrator stars shows the $2.2056\,\mu$m \ion{Na}{I} line
in absorption,
but this line is very narrow (much less
broad than the other lines), suggesting an interstellar origin for the absorption.

\subsubsection{CO band heads}
The CO ro-vibrational bands, with their band heads at
$2.2935\,\mu$m for the $v = 2 - 0$ transitions, 
$2.3227\,\mu$m for the $v = 3 -1$ transitions, and
$2.3535\,\mu$m for the $v = 4 -2$ transitions,
constitute very prominent absorption features
in cool stars. They are generally seen at spectral types later than mid G.
None of our OB2 calibrator spectra show these band heads.

\subsection{Spectral type determination for the O- and early B-type stars \label{spectral-typing.sec}}

Spectral classification of early-type stars in the infrared regime
is less reliable than in the optical \citep[see][]{Gray},
mostly because the lines
in the NIR regime are often formed in higher atmospheric layers
than the optical lines.
In particular, the equivalent width of the H and He lines in hot stars
not only  depends on the effective temperature, but is also strongly
influenced by the surface gravity and the detailed properties of the
(generally very strong) stellar winds 
\citep[see, e.g.,][]{Hanson05,2005A&A...440..261R}.
Nevertheless, the He lines show a systematic
increase or decrease in line strength with spectral type,
which allows spectral types to be reasonably well determined.

The general trends of decreasing line strengths with increasing
temperature for the H and \ion{He}{I} lines and the
increasing line strengths with increasing
temperature for the \ion{He}{II} lines are clearly seen in the
spectra of our calibrator stars, although with a substantial
scatter.
Given the moderate spectral resolution of our spectra and the
limited number of just four diagnostic He lines we can use,
we do not aim at a very detailed and highly accurate spectra modeling but instead restrict our analysis to
estimating spectral types with an accuracy of about two subtypes.

For all the stars in our samples that showed clear He lines, we analyzed the
strengths of the diagnostic lines described below
to determine their spectral type as accurately as possible.
In order to avoid biases in the classification procedure,
we did our classification for all KMOS spectra (i.e.,~including those stars
for which a spectral type was already known from the literature). For stars that were
observed twice, the classification was also done twice. The analysis was also done
in a ``blind'' way, that is, without taking the literature information into account.
Only after finishing the classification did we check the corresponding literature
values, and we compared our two independent spectral types for those stars
that were observed twice. We generally found rather good agreement (within about
two subtypes) in all cases.

\subsection{Excluding the presence of Red Supergiants \label{RSG.sec}}

As described above, we consider cool stars to be either foreground field stars
or background giants, in any case unrelated to the Carina Nebula.
However,
a potential complication arises from the fact that some evolved massive stars
spend a small fraction of time near the end of their lifetimes as red supergiants (RSGs) and display (very) cool
spectra in this phase. Young stellar populations with ages of at least about 10~Myr
can thus contain RSGs. Although it appears likely that the stars in Tr~16-SE
are too young to be RSGs (the oldest age of the well-studied clusters 
in the Carina Nebula is~8~Myr [for Tr~15]; see \citet{CCCP-HAWKI}), this possibility 
should not be a priori excluded.
Therefore, we
checked whether the cool stars might possibly be RSGs in the Carina Nebula.

For this, we used the fact that 
RSGs generally have bolometric luminosities of $\ga 10^4\;L_\odot$ and
employed the magnitude-luminosity calibration for RSGs derived by
\citet{2013ApJ...767....3D}.
Although the (unknown) extinction toward our target stars is a possible complication, 
it is not a very serious one
for the purposes of our estimates for two reasons: First, we work in the infrared, and the
extinction in the $K$-band is almost a factor of ten lower than in the optical.
Second, from the color-color diagram of our target stars as well as from the cloud column density maps from our previous APEX/LABOCA submillimeter and 
\textit{Herschel} far-infrared
maps, we know that the cloud extinction in TR~16-SE ranges from values of just a few
magnitudes to about 15 magnitudes in optical extinction, corresponding to $A_K$ values
of $\la 1.5$~mag.
Since the minimal luminosity of RSGs is $\approx 10^4\;L_\odot$,
this implies that, even for an extinction of $A_V \approx 15$~mag,
all possible RSGs should be brighter than $K \approx 5.7$.
We note that this value is consistent with corresponding estimates
based on the $K$-band bolometric corrections for cool stars from \citet{2010MNRAS.403.1592B};
thus, this magnitude limit is robust.

All our target stars have $K$-band magnitudes $>6.74$ (and $> 7.7$ for all but the two brightest stars).
We can therefore conclude that even the brightest stars in our sample
are substantially fainter (by more than one magnitude in the $K$-band) than the 
expected magnitudes of even the faintest RSGs in the Carina Nebula.
The cool stars in our sample must therefore be either foreground stars
or giants (and perhaps RSGs) in the galactic background.

\subsection{Results of the spectral classification} \label{results}

Our spectral typing results for each target star are 
listed in Table \ref{results.table}.
The statistics as a function of spectral types are as follows.

Three stars are classified as "early O-type stars" ($\le$ O5):
  B8 (O5-6), B13 (O5), and B14 (O5).
One of these, B13 = V662~Car = MJ~596, was known before our observation 
to be a double-lined O5.5V+O9.5V spectroscopic binary \citep{Niemela06} and was
reclassified to a spectral type of  O5Vz+B0:V by \citet{2016AJ....152..190A}.
\smallskip

Three stars (B5, F3, and B7) are classified as "mid O-type stars" (O6 -- O7).
We note that for our target star, B5, an optically determined spectral type  of O6~V
became available in the literature  \citep{2016AJ....152..190A} after our
KMOS observations.
\smallskip

One star (F8) is classified as a "late O-type star" (O8 -- O9), five stars (F7, B12, B18, F22, and B21)
as "late O- to early B-type stars" (O9 -- B1), and 
three stars (F9, F16, and F17)  as "early B-type stars" (B0 -- B2).
Eight stars (F10,  F11, F12-2, B11, F14, B16, B17, and F21)
are classified as "likely AeBe stars" (i.e., intermediate-mass 
pre-main-sequence stars).
\smallskip

The total combined number of OB2 and likely AeBe stars in the Tr~16-SE region is thus $23$ ($15$ plus eight).
 We note that one of the three target stars outside Tr~16-SE, C17 in Tr~16, is also
classified as a likely AeBe star.
\medskip

The spectra of 18 stars in Tr~16-SE showed no indications of He lines
but clear signatures of cool photospheres
(e.g., CO absorption and/or Na and Ca line absorption), and
are thus classified as cool stars.
Six stars (F4, F12-1, F13, F15, B22, and B23) could not be reliably classified;
some of them may be late B- or A-type stars.

\section{Comparison to previous OB-candidate samples and photometric spectral type estimates \label{Candidates.sec}}

\citet{Povich11a} used photometry from 2MASS and \textit{Spitzer}
to analyze the spectral energy distributions (SEDs) of infrared sources in the
Carina Nebula. In Tr~16-SE, they identified six stars (their OBc 48, 50, 51, 52, 56, and 61)
as candidate OB stars based on their observed SED.
All six of these stars are in our sample, and 
all of them are classified as O-type or early B-type stars
in our independent spectroscopic analysis.
This shows that the SED-based classification of OB candidate stars
performed by \citet{Povich11a} worked very well and yielded reliable results.

In addition to the six OB candidates from \citet{Povich11a}, our spectroscopic survey 
yields a further nine OB2 stars that were not in their OB-candidate sample.
One of these is the previously known O5 star MJ~596, and our analysis (see below) 
suggests that six of these
additional OB2 stars are background objects. However, two of these additional OB2 stars
(B12 and B18, both classified as O9--B1 stars) are found to be 
members of Tr~16-SE.
This shows that, as expected, the sample of SED-selected OB candidates 
from \citet{Povich11a} is not 100\% complete;
considering our results, the level of completeness is
estimated to be  $6/8 = 75\%$.

For several of our target stars, \citet{2017MNRAS.465.1807M} reported spectral 
type estimates based on an analysis of their optical-infrared SED\footnote{We note that \citet{2017MNRAS.465.1807M} also obtained optical 
spectra of some of their target stars,
but none of these spectrally observed stars are located in the area of our 
TR~16-SE KMOS observations.}
 after our KMOS observations.
We compared their spectral type estimate to our KMOS spectroscopic classifications
and found generally rather good agreement, typically within one or two subtypes.

\section{Clues on stellar distances from Gaia data \label{Gaia.sec}}

As mentioned in Sect.~\ref{strategy.sec}, our target stars were selected based only on their
position on the sky and a limit on their $K_s$-band magnitude. 
Our target sample therefore contains not only stars in the Carina Nebula, but also
some foreground and background stars.
The Gaia mission \citep{Gaia} has recently determined
parallax values for many of our target stars, providing very important distance information (which was not
available at the time of our observations).
We therefore checked whether we can use the Gaia data to identify 
foreground or background stars in our sample.
This requires a good
knowledge of the distance to the Carina Nebula and Tr\,16-SE.

\subsection{The distance to the Carina Nebula}

As described in \citet{SB08}, the pre-Gaia distance to the central region of the
Carina Nebula is well known to be 2.3~kpc. Since the projected extent of 
the whole Carina Nebula cloud complex is about 100~pc \citep{Preibisch12,2017A&A...605A..85P}, 
different parts of the complex
may well be about 100~pc closer or more distant.

Although the Gaia data offer a good opportunity
to investigate the distances in a new and independent way,
the typical uncertainties
of the Gaia Early Data Release 3 (EDR3) parallaxes and proper motions for most stars
in the Carina Nebula are still too large
to allow a purely astrometric identification of the Carina Nebula association members.
One therefore needs a preselected sample of member candidates to determine
distances with the Gaia data. Using the Gaia DR2 data,
\citet{Povich19} presented a determination of the distance to the 
large sample of $\,> 10\,000$ X-ray selected
stars in the Carina Nebula from the CCCP project \citep[see][]{2011ApJS..194....4B}
and found a value of $2.50^{+0.28}_{-0.23}$~kpc,
slightly higher than but still consistent with the previous ``canonical'' value of 2.3~kpc.

Taking advantage of the recently released Gaia EDR3 data \citep[see][]{Gaia-EDR3},
which provide improved accuracy of the astrometric parameters,
we determined the mean value of the parallaxes of
the 10\,714 X-ray selected ``likely Carina member stars''
from the CCCP project; this yielded 
 $\langle \varpi \rangle = 0.4347$~mas, corresponding to a distance 
of $\approx 2.30$~kpc, in very good agreement with the pre-Gaia distance value.

\subsection{Gaia distances of the KMOS target stars}

For 58 of the stars observed with KMOS,
we could find a 
Gaia EDR3 match within a search radius of one arcsecond.
For six stars, the listed parallaxes
are negative (i.e.,~provide no useful information about the distance).
The uncertainties of the Gaia parallaxes of our KMOS target stars range from
$\sigma_\varpi = 0.0102$~mas up to $\sigma_\varpi = 0.5119$~mas.

Assuming that the stars in the Carina Nebula are in the distance range
between $\approx 2.2$~kpc and $\approx 2.8$~kpc, we can check
how consistent the parallax of individual KMOS target stars 
is with this  distance range.
For the large majority of our KMOS target stars, 
the $\left[\varpi  \pm 3 \sigma_\varpi \right]$ 
interval overlaps with the  $\left[ \frac{1}{2.8} -  \frac{1}{2.2} \right]$~mas range; 
we consider
the parallaxes of these stars as being consistent with membership in the Carina Nebula.

\subsection{Foreground stars}

We consider stars as "likely foreground stars" if their $3 \sigma$ lower bound
of the measured parallax is $ \varpi - 3 \sigma_\varpi > \frac{1}{2.2}$~mas, 
and as
"clear foreground stars" if their $5 \sigma$ lower bound
of the measured parallax is $ \varpi - 5 \sigma_\varpi > \frac{1}{2.2}$~mas. 
According to these definitions, five of our target stars are clear foreground stars:
B3=C21 ($1/\varpi = 1442$~pc), 
B4=C22 ($1/\varpi = 389$~pc),
B22 ($1/\varpi = 486$~pc),
B23 ($1/\varpi = 1398$~pc),
and 
F24 ($1/\varpi = 1337$~pc).
In addition, the two target stars
C18 ($1/\varpi = 1991$~pc)
and
F23 ($1/\varpi = 1934$~pc)
are likely foreground stars.

In our KMOS spectral classification, the
 Gaia-based foreground stars 
B3, F24, and C18  were classified as cool stars,
B4 as a spectral type between F and G, and
B22 and B23 as possible B- or A-type stars;
these spectral types are consistent with their nature as
field stars in the foreground. 
None of the OB2 stars in our KMOS sample have Gaia parallaxes
that would suggest their being foreground stars.

\subsection{Background stars}
A similar attempt to identify possible background stars was not successful.
Although a couple of stars have suspiciously small parallaxes
($\varpi < 0.2$~mas), the $5\sigma$ interval
still overlaps with the $[2.2 - 2.8]$~kpc Carina Nebula distance range for all stars.
Therefore, none of our target stars can be reliably classified as background stars
based on their Gaia parallaxes. 
We note that it is very likely that
some of our target stars actually are in the background (e.g., as red giants), 
but the accuracy of the current Gaia parallaxes is just not (yet) good enough 
to prove this directly from their parallaxes. 

\section{Properties of the OB2 stars \label{OB2-stars.sect}}

\setcounter{table}{3}
\begin{table*}
\caption{Properties of the OB2 stars in the Tr~16-SE region.} \label{tab_OB2stars}
\begin{tabular}{lllrclrr}
\hline\hline                 
Star Name / Position & Spectrum & \multicolumn{1}{l}{Spectral} & \multicolumn{1}{c}{$A_V$} & $\log \left(L_{\rm bol}/L_\odot\right)$ for & $\log \left(L_{\rm bol}/L_\odot\right)$ & Membership & Comment \\
\multicolumn{1}{c}{ J2000 } &  & \multicolumn{1}{c}{Type} & \multicolumn{1}{c}{[mag]} & $D=2.3$~kpc
 & expected  &  &\\
\hline\noalign{\smallskip}
\hline                                    
J10453631--5948233 & B13   & {O5}    &  6.4  & 5.33 &5.49&  Tr~16-SE member & MJ~596 \\ 
J10453674--5947020 & B14   & {O5}    & 10.3  & 5.57 &5.49 &  Tr~16-SE member & \\      
J10453024--5948206 & B8    & {O5--6} & 11.8  & 5.50 &5.32--5.49 & Tr~16-SE member &\\ 
J10452227--5950470 & B5    & {O6}    &  6.4  & 5.58 &5.32 & Tr~16-SE member & MJ~568 \\ 
J10452862--5947553 & B7+C24& {O6--7} & 14.5  & 5.21 &5.14--5.32 & Tr~16-SE member & \\ 
J10453470--5947537 & B12   & {O9--B1}& 13.3  & 4.76 &4.37--4.77 & Tr~16-SE member & \\ 
J10454460--5950411 & B18   & {O9--B1}&  5.4  & 4.75 &4.37--4.77&  Tr~16-SE member &\\ 
J10454661--5948404 & B21   & {O9--B1}&  7.7  & 4.25 &4.37--4.77 & Tr~16-SE member &\\ 
J10451717--5947013 & F7+C20& {O9--B1}&  8.8  & 4.27 &4.37--4.77 & Tr\,16-SE member & \\ \hline
J10450879--5950537 & F3    & {O6--7} & 10.5  & 4.27 &5.14--5.23 &  Background star &\\
J10452013--5950104 & F8    & {O8--9} &  7.4  & 3.91 &4.77--4.96 &  Background star &\\
J10454595--5949075 & F22   & {O9--B1}&  5.8  & 3.83 &4.37--4.77&   Background star &\\ 
J10453755--5948529 & F16   & {B0--1} &  5.5  & 3.59 &4.37--4.57 &  Background star &\\
J10453774--5947552 & F17   & {B0--1} &  9.1  & 3.86 &4.37--4.57 &  Background star &\\
J10452648--5946188 & F9+C23& {B1--2} &  7.2  & 3.84 &4.19--4.37 &  Background star & \\
\hline                                   
\end{tabular}
\end{table*}

Since the Gaia data cannot reliably discern stars in TR~16-SE
from objects in the galactic background,
we need an alternative way to identify possible background stars
in our OB2 sample.
We therefore calculated the extinction, the absolute $K$-band magnitude,
and, from this, the bolometric magnitude for an assumed distance
of 2.3~kpc. The corresponding luminosity estimate is then compared
to the expected luminosity of main-sequence stars of this 
spectral type; if both values agree within a factor of two,
we assume the observed stellar properties to be consistent
with membership in the Carina Nebula.

We determined the extinction from the $H-K_s$ color excess.
For the O-type stars, we 
assumed the intrinsic colors listed in 
\citep{2005A&A...436.1049M} and then used
the bolometric corrections $BC_K$ from \citet{2006A&A...457..637M}
to calculate absolute magnitudes and luminosities for $D = 2.3$~kpc.
For the early B-type stars, we used the intrinsic colors and bolometric
corrections from \citet{2013ApJS..208....9P}.
The calculated luminosities $L[2.3\,{\rm kpc}]$ were then compared to the luminosities
listed for the corresponding spectral types in \citet{Smith06}.

We found $K$-band extinction values between 0.64~mag and 1.60~mag,
corresponding to
visual extinction values  of $A_V \sim [5.8 - 14.5]$~mag
for the O-type stars in Tr~16-SE.
For the O-type stars in Tr~16, which we observed as spectral calibrators, 
a similar analysis yielded 
$K$-band extinction values between 0.24~mag and 0.44~mag, corresponding to
visual extinctions of $A_V \sim  [2.2 - 4]$~mag.

For nine of the 15 OB2 stars in Tr~16-SE, our luminosity estimates are consistent with their being stars in the Carina Nebula complex, and these stars are thus denoted as ``Tr~16-SE members''
in Table \ref{tab_OB2stars}.
However, for three O-type stars and three early B-type stars, 
the computed bolometric luminosity
for $D = 2.3$~kpc is more than a factor of 2.5 lower than the expected
 luminosity for main-sequence stars of this spectral type; these are
denoted as ``background stars'' 
in Table \ref{tab_OB2stars}.

   \begin{figure}
\includegraphics[width=8.5cm,trim={30 140 65 170},clip]{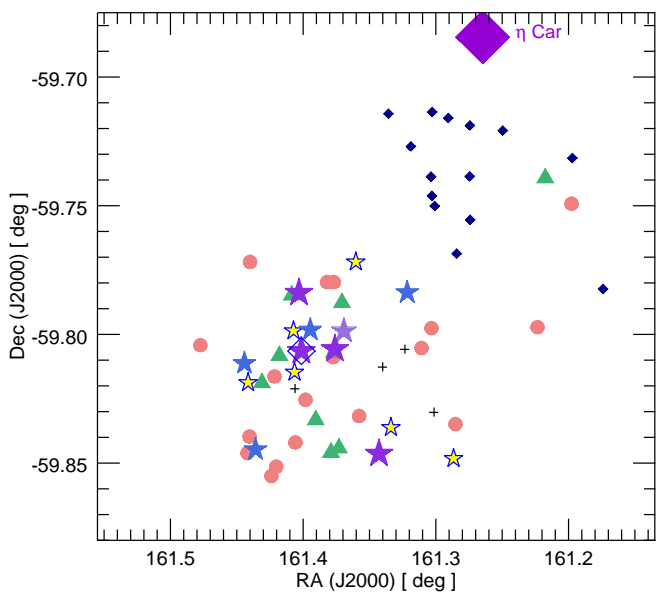}
      \caption{Map showing the spatial distribution of our
KMOS target stars in celestial coordinates.
O-type stars are denoted by violet star symbols,
early B-type stars with light blue star symbols,
(Ae)Be stars with green triangles,
and late-type stars with light red circles.
The previously known O-type star MJ~596 near the center of the cluster
is denoted by an empty diamond (superimposed onto the star symbol).
The OB2 stars that we classified as background stars
are shown with yellow-filled star symbols.
Stars that could not be classified
are shown as small crosses, and
the OB calibrator stars in Tr~16 are shown with small dark blue
diamonds.
The position of $\eta$~Car at the top edge of the map is also indicated.
}
         \label{Fig:map}
   \end{figure}
%

\section{The stellar population of Tr~16-SE}

In Fig.~\ref{Fig:map} we show the spatial distribution
and spectral classifications of our KMOS target stars.
We find a clear concentration of OB2 cluster members
near the center of the cluster, which is located roughly at the
position of the O5 star MJ~596.
\medskip

%
   \begin{figure}
\includegraphics[width=8.3cm,trim=40 140 65 170,clip]{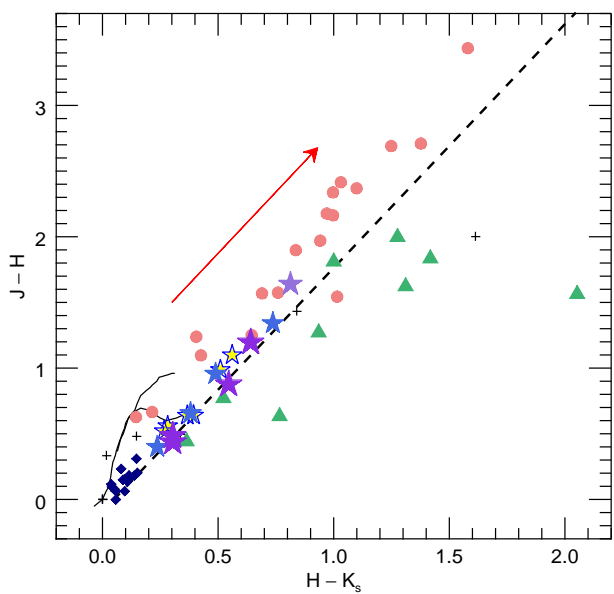}
      \caption{Near-infrared color-color diagram of the
KMOS target stars. 
The solid lines show the color of the main-sequence and giant stars,
the arrow shows an $A_V = 10$ mag reddening vector with slope 1.85 \citep[see][]{VISTA2},
and the thick dashed line marks the separation between the photospheric reddening
band (above this line) and the infrared excess region (below this line).
Symbols and colors for the stars are the same as in Fig.~\ref{Fig:map}.
}
         \label{Fig:ccd}
   \end{figure}
%

In Fig.~\ref{Fig:ccd} we show the NIR color-color diagram
of the KMOS target stars.
It shows that
all OB2 stars show extinction values of more than 
$A_V \ge 5$~mag, clearly suggesting them to be located
in or (more likely) behind the clouds that obscure Tr~16-SE, as expected.
While none of the OB2 stars show evidence for an infrared excess,
we note that many (Ae)Be stars do clearly show infrared excesses,
as expected for such young stellar objects.
Most of the cool stars show high extinction values ($A_V \ge 10$~mag),
as expected for background objects that are seen through the clouds
obscuring Tr~16-SE.
\medskip

%
   \begin{figure}
\includegraphics[width=8.2cm,trim=70 85 85 115,clip]{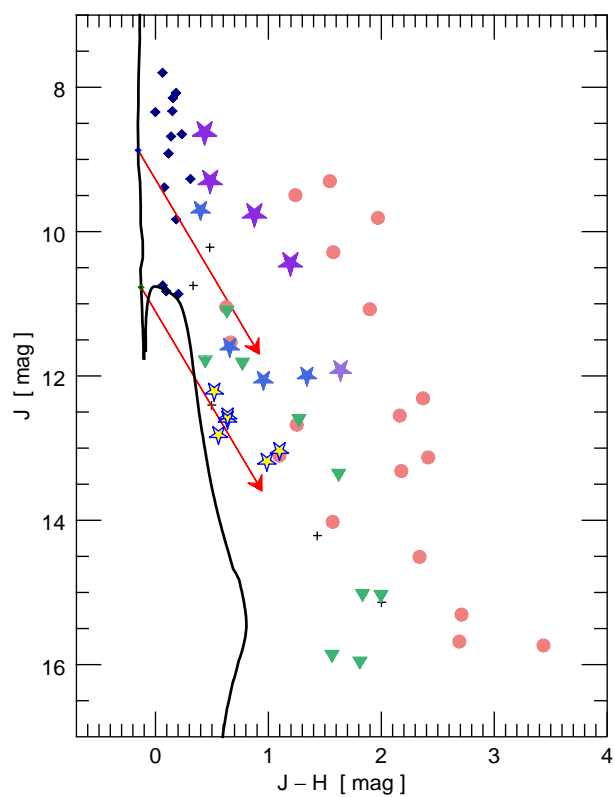}
      \caption{$J$ versus $J-H$ color-magnitude diagram
of the
KMOS target stars. 
The solid black line shows a 1~Myr isochrone based on the 
PARSEC stellar models \citep[see][]{2012MNRAS.427..127B}.
The arrows indicate reddening vectors for $A_V = 10$~mag starting at the
location of 10~Myr old stars with masses of $20\,M_\odot$ and
$8\,M_\odot$.
Symbols and colors for the stars are the same as in Fig.~\ref{Fig:map}.
}
         \label{Fig:JHcmd}
   \end{figure}
%

In Fig.~\ref{Fig:JHcmd} we show an NIR color-magnitude diagram
of the KMOS target stars.
The locations of the different groups
in this diagram are consistent with their spectral
classification, which also suggests that a large fraction
of the cool stars in our KMOS sample must be red giant stars
in the galactic background.

   \begin{figure*}
   \centering
\includegraphics[width=15cm,]{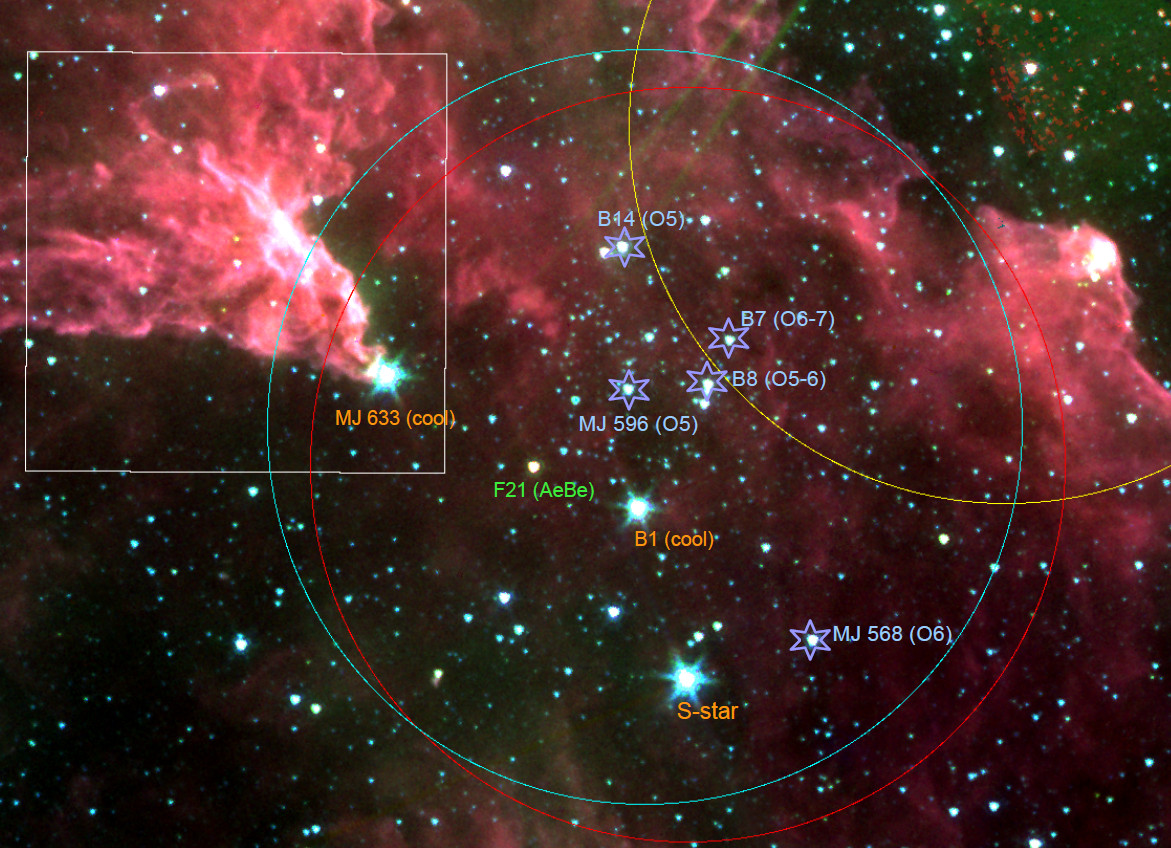}
\caption{\textit{Spitzer} Infrared Array Camera (IRAC) color-composite image of Tr~16-SE 
and the pillar G~287.75$-0.66$; 
the image is composed of the $3.6\,\mu$m image in the blue color channel,
the $4.2\,\mu$m image in the green color channel, and the
$5.8\,\mu$m image in the red color channel. North is up, and east is to the left.
The white $4' \times 4'$ square marks the region of the pillar G~287.75$-0.66$, and
the circles show the KMOS patrol fields for the \textbf{B} (blue), \textbf{F} (red),
and \textbf{C }(yellow) observations.
North is up, and east is to the left.
The early O-type stars in Tr~16-SE are marked by the empty violet star symbols;
their names and spectral types are given in the annotation.
Furthermore, some of the brightest infrared sources 
(including the S-type star 2MASS~J10453185-5951094) are annotated.
}
         \label{Fig:pillar-Tr16SE}
   \end{figure*}
%

\section{Tr~16-SE and the pillar G~287.75--0.66}

The \textit{Spitzer} image in Fig.~\ref{Fig:pillar-Tr16SE}
 shows the spatial relation between the pillar G~287.75$-0.66$ and the stars in Tr~16-SE.
It shows a bright infrared source
located directly at the tip of the pillar. 
Optical images also show a bright ($V = 13.14$) star (MJ~633)
at the pillar tip.  This star may thus appear as
the ``prime suspect'' for the radiation source creating and
shaping the pillar. 
However, a detailed inspection of optical\footnote{
We used a 25 sec exposure r\texttt{\_}SDSS-band image obtained
with the ESO VLT Survey Telescope (file name OMEGA.2012-02-15T05:34:43.880.fits) that was obtained
as part of the \textit{VST Photometric H-$\alpha$ Survey
of the Southern Galactic Plane} \citep[see][]{Drew14}; it has  an image scale of $0.21\arcsec$ per pixel
and the FWHM of the PSF is
$\approx 0.8\arcsec$.}
and infrared\footnote{We used our 720 sec exposure $J$-band image obtained with HAWK-I at the
ESO VLT \citep[see][]{HAWKI-survey}, which provides an angular resolution of
$\approx 0.5\arcsec$.}
 images
with subarcsecond angular resolution clearly shows that the bright infrared source does not coincide with the optical star MJ~633: There are two distinct sources with an angular separation of $1.8\arcsec$.
 The star MJ~633 is seen as a bright object in the optical image
and as a moderately bright object in the NIR
image at a consistent position of $\alpha ({\rm J2000}) = 10^{\rm h}\,45^{\rm m}\,54.807^{\rm s}$,
 $\delta ({\rm J2000}) = -59\degr\,48\arcmin\,15.56\arcsec$. 

The position of the bright infrared source measured in our HAWK-I image
is $\alpha ({\rm J2000}) = 10^{\rm h}\,45^{\rm m}\,54.575^{\rm s}$,
$\delta ({\rm J2000}) =  -59\degr\,48\arcmin\,15.1\arcsec$; it
can be identified with the 2MASS point source
J10455458-5948150.
This object is also a bright
mid-infrared source detected by \citet{Mottram07} and
classified as a massive
young stellar object candidate with the designation G~287.7441$-00.6869$~1.
In the optical image, the infrared source is seen as a very faint 
brightness enhancement in the wings of the PSF of the
(optically) much brighter star MJ~633.
\medskip

In our KMOS observations we obtained spectra of both stars.
Our KMOS spectra of the
optically visible star MJ~633 (= target F24)
show no \ion{He}{I} or \ion{He}{II} lines,
but do show clear CO band head absorption as well as Na and Ca absorption lines.
This clearly suggests that MJ~633
is a cool star. Furthermore, the Gaia EDR3 parallax of $\varpi = 0.7480 \pm 0.0127$~mag
corresponds to a distance of $ D= 1.337\; [1.314 - 1.360] $~kpc and 
shows that this star is a foreground object, unrelated to the 
Carina Nebula. It is just a random projection effect that puts this star
apparently at the tip of the pillar.

Our KMOS spectra of the optically invisible infrared-bright star J10455458-5948150 (i.e., target B24)
also show no clear signs of \ion{He}{I} or \ion{He}{II} lines, but do show
CO band head absorption as well as Na and Ca absorption, clearly suggesting
a cool photosphere.
We conclude that the previous classification of this source as a massive young stellar object candidate
by \citet{Mottram07} is not confirmed by our spectra.
However, since our spectrum shows the Bracket $\gamma$ line to be in emission,
the object may well be a young stellar object, though not a massive one.

In conclusion, the two stars seen apparently at the tip of the pillar are either unrelated to the Carina Nebula
or are not luminous enough to be responsible for producing the pillar.
The visual impression that one of these stars is the irradiating source of the
pillar is purely a random projection effect.
\medskip

In Fig.~\ref{Fig:pillar-Tr16SE} we mark and annotate the early O-type 
stars in Tr~16-SE;
as the emitted stellar UV luminosity is steeply increasing with effective temperature,
we only consider the five stars in Tr~16-SE with spectral types between O5 and O6.
These five O5-O6 stars are located at projected distances between $\approx 1.7$~pc and 
$\approx 3.1$~pc from the pillar; their combined UV irradiation may have shaped
 the pillar G~287.75$-0.66$.

\section{Conclusions and summary}

Our KMOS observations of a magnitude-limited sample
of possible OB stars in the obscured cluster Tr~16-SE
led to the identification of  five stars with spectral types between O5 and O7 (only two of which
were known before)
and four new stars with spectral types between O9 and B1
as members of Tr~16-SE.
With at least five O-type stars and a further four late O- or early B-type stars,
Tr~16-SE is one of the 
larger clusterings of high-mass stars in the Carina Nebula;
its high-mass stellar population is
smaller than the  prominent open clusters
Tr~14, 15, and 16 (each of which contains $\ge 15$ OB2 stars) 
but is comparable to or larger than the clusters Bochum~10 (15 OB2 stars) and
Bochum 11 (seven OB2 stars).
Tr~16-SE therefore constitutes a significant, but so far overlooked, part
of the massive cluster population in the Carina Nebula complex.

 Our newly identified O-type stars in Tr~16-SE significantly increase the census of
spectroscopically identified O-type stars in the Carina Nebula (previously 70 stars).
Another very important result is the fact that all\ six of the OB candidates 
in Tr~16-SE from 
\citet{Povich11a} are spectroscopically confirmed as OB2 stars by our KMOS observations. 
This suggests that the large majority of their 94 OB candidates 
in the full area of the Carina Nebula are most likely true OB2 stars.

Assuming a $\approx 75\%$ completeness for the
\citet{Povich11a} OB-candidate sample (see Sect.~\ref{Candidates.sec}),
this suggests that 
that the Carina Nebula probably contains $\approx (94-6) / 0.75 \approx 117$
further OB2 stars awaiting spectroscopic confirmation.
The total OB2 star population in the Carina Nebula may well be twice
as large as the presently identified sample.

\begin{acknowledgements}
     We thank  Rolf Kudritzki and Joachim Puls for enlightening discussion
about the spectra of massive stars. 
The LMU Bachelor Physics student D.~Beckord provided assistance 
in the construction of the initial spectral line catalog.
The research of T.P.~and B.E.~was partly supported by the Excellence Cluster ORIGINS 
which is funded by the 
Deutsche Forschungsgemeinschaft (DFG, German Research Foundation) under 
Germany's Excellence Strategy -  EXC-2094 - 390783311.
This work has made use of data from the European Space Agency (ESA) mission
{\it Gaia} (\url{https://www.cosmos.esa.int/gaia}), processed by the {\it Gaia}
Data Processing and Analysis Consortium (DPAC,
\url{https://www.cosmos.esa.int/web/gaia/dpac/consortium}). Funding for the DPAC
has been provided by national institutions, in particular the institutions
participating in the {\it Gaia} Multilateral Agreement.
\end{acknowledgements}

%
%

\bibliographystyle{aa}
\bibliography{ref}

\begin{appendix} 

\section{Spectra}

\begin{figure*}
 \centering
 \includegraphics*[width=15.8cm,viewport=50 82 526 818]{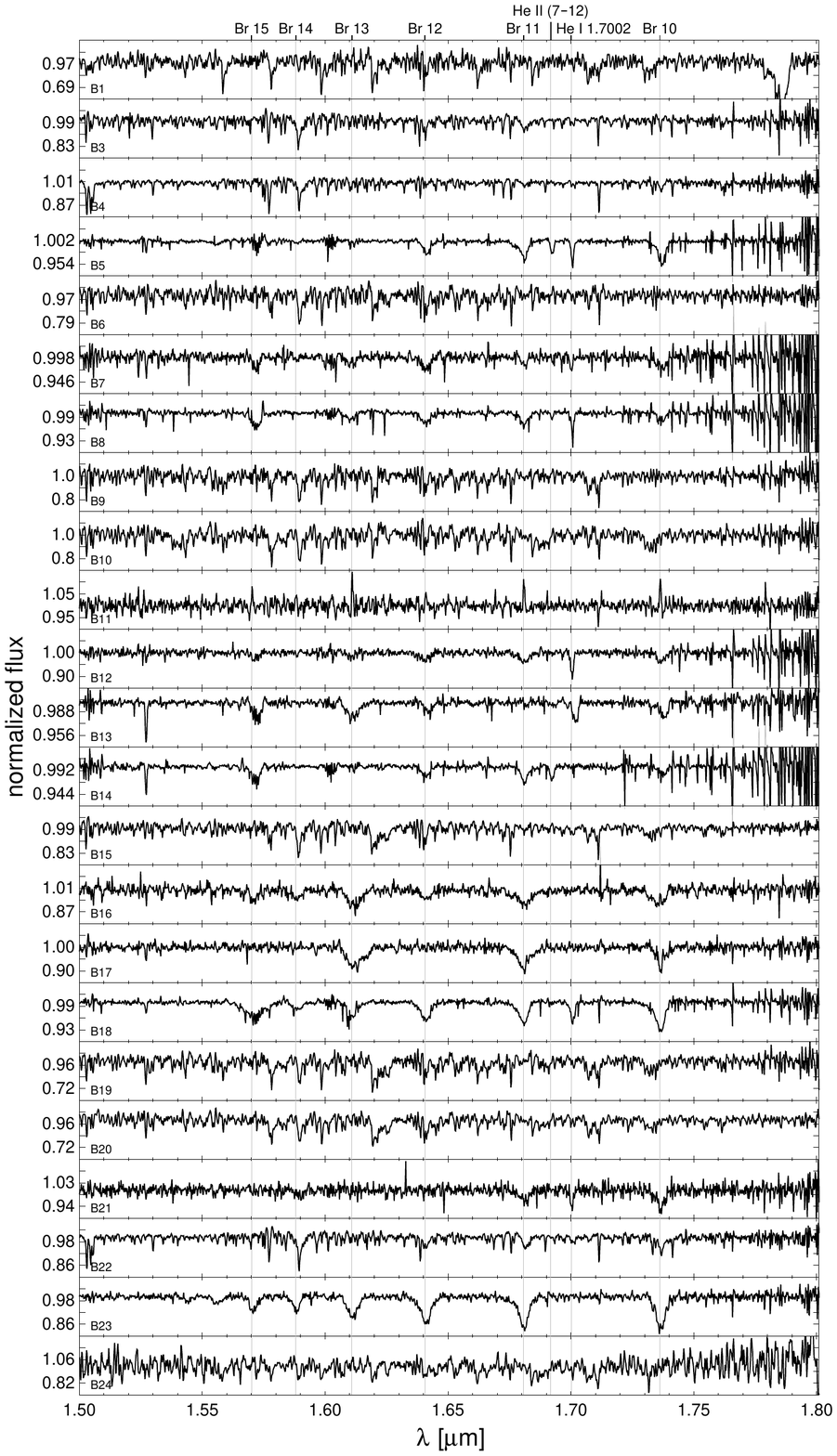}
      \caption{$H$-band spectra in the\textbf{ B} sample.}
         \label{fig:spectra-B-H}
\end{figure*}

\begin{figure*}
 \centering
 \includegraphics*[width=15.8cm,viewport=50 82 526 818]{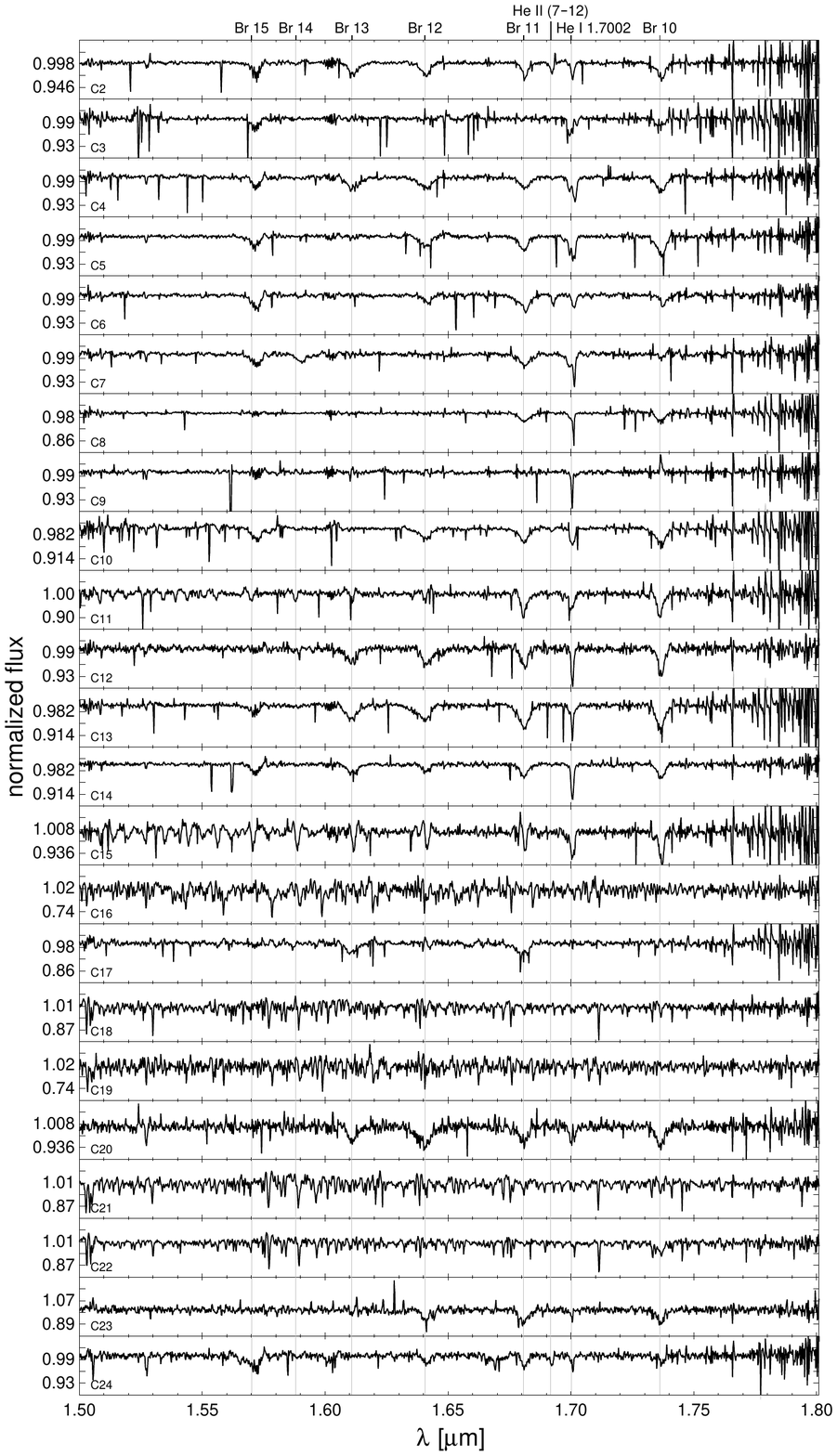}
      \caption{$H$-band spectra in the \textbf{C} sample.}
         \label{fig:spectra-C-H}
\end{figure*}

\begin{figure*}
 \centering
 \includegraphics*[width=15.8cm,viewport=50 82 526 818]{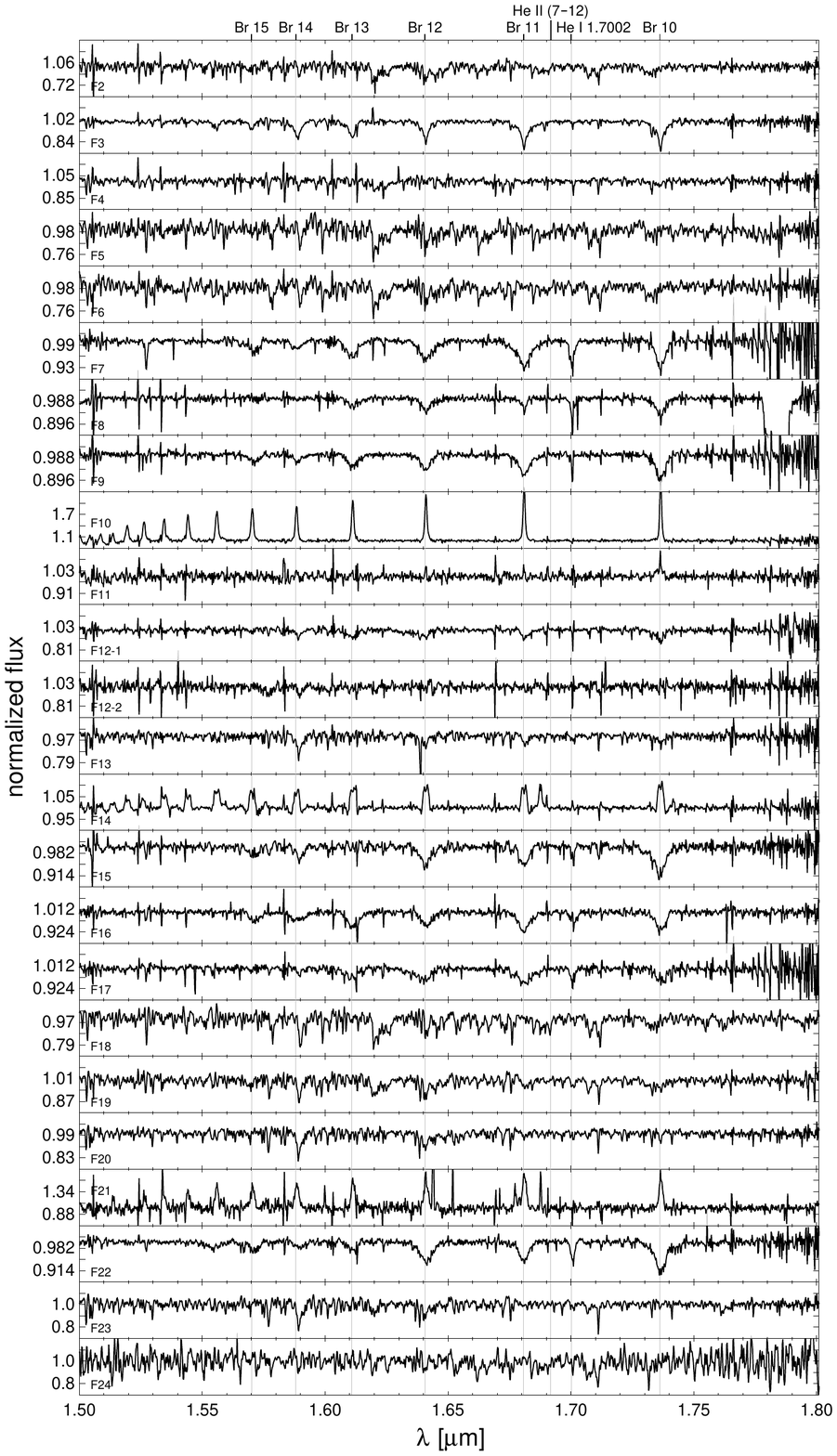}
      \caption{$H$-band spectra in the \textbf{F} sample.}
         \label{fig:spectra-F-H}
\end{figure*}

\begin{figure*}
 \centering
 \includegraphics*[width=15.8cm,viewport=50 82 526 818]{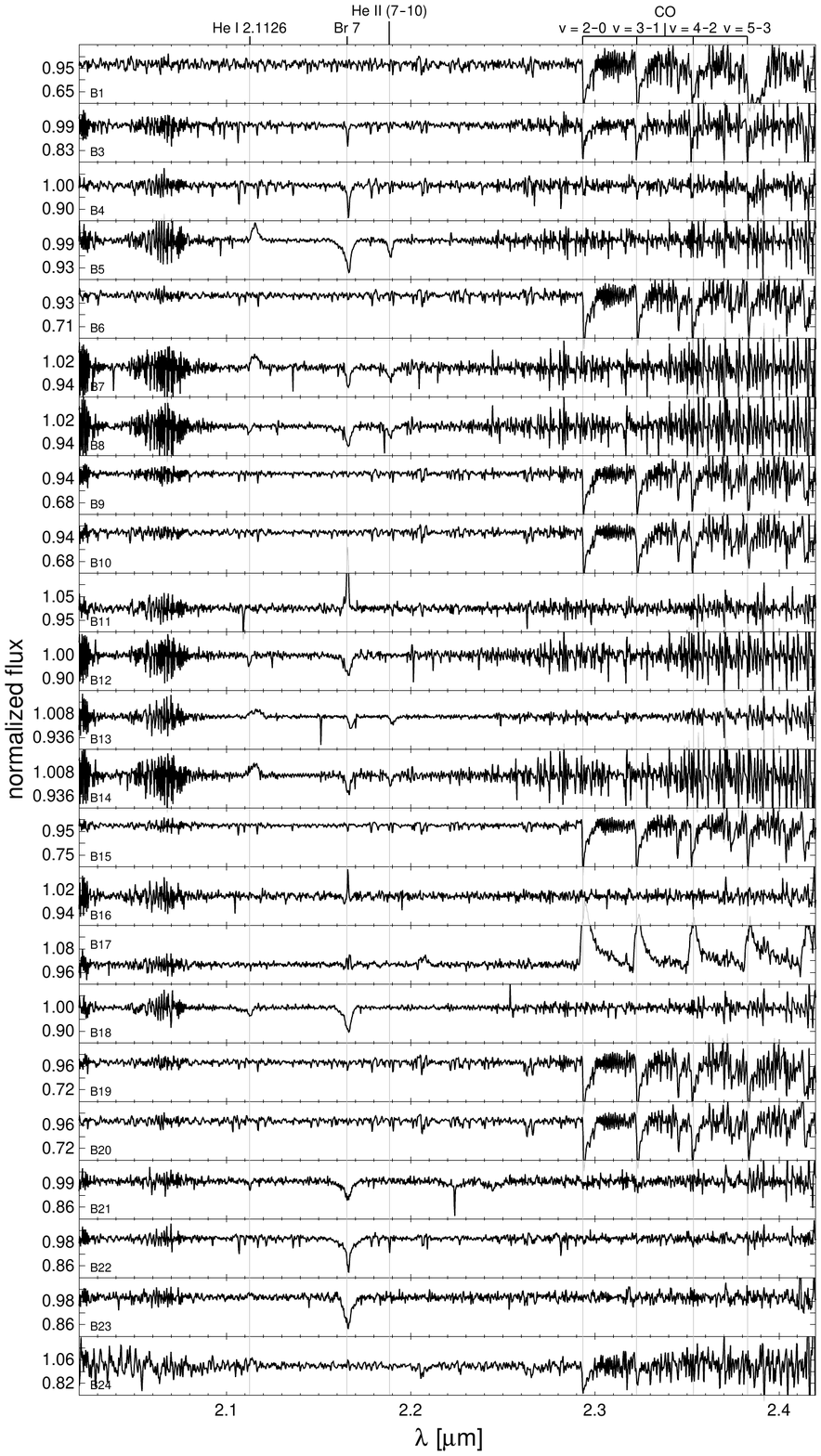}
      \caption{$K$-band spectra in the \textbf{B} sample.}
         \label{fig:spectra-B-K}
\end{figure*}

\begin{figure*}
 \centering
 \includegraphics*[width=15.8cm,viewport=50 82 526 818]{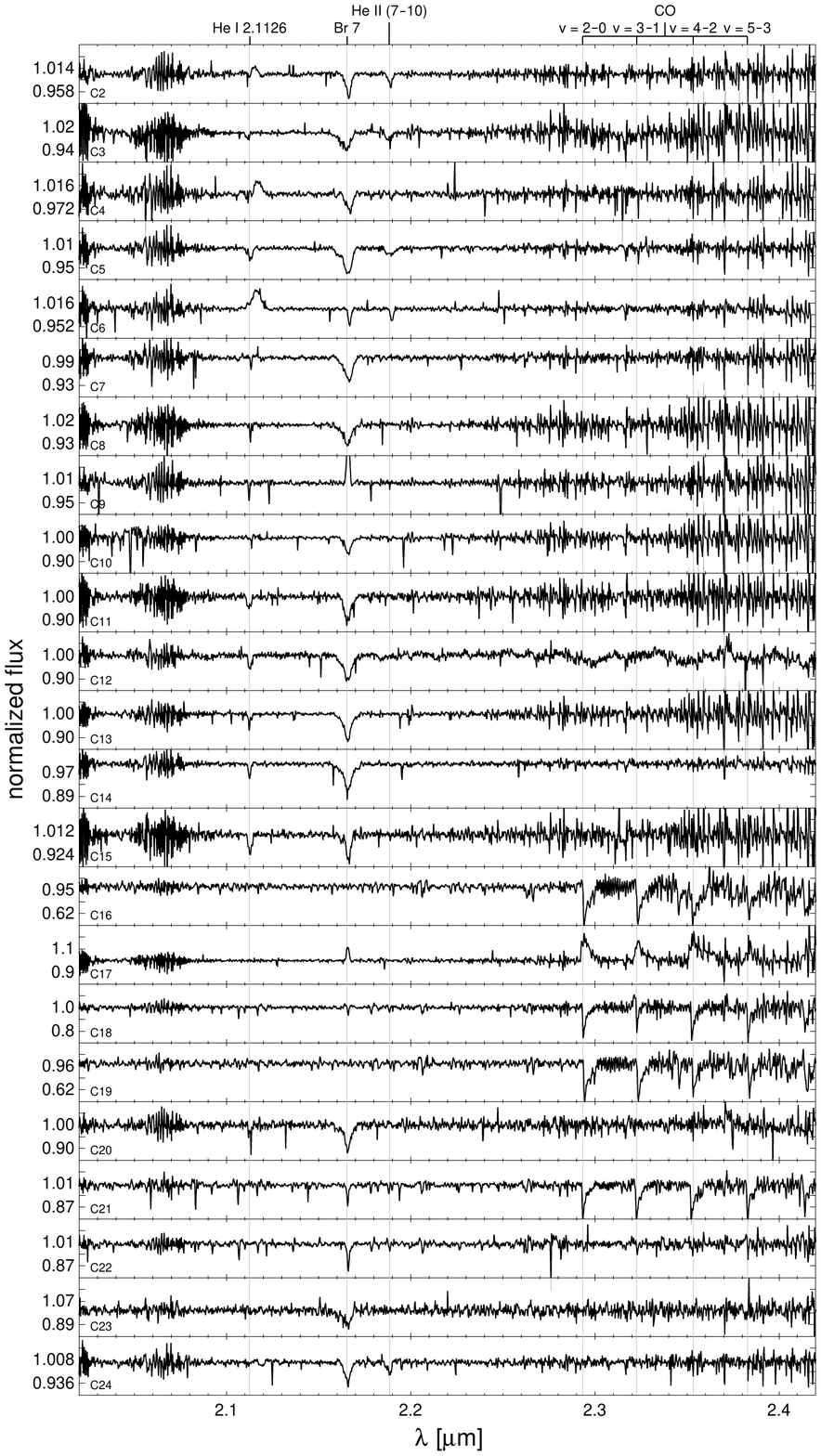}
      \caption{$K$-band spectra in the \textbf{C} sample.}
         \label{fig:spectra-C-K}
\end{figure*}

\begin{figure*}
 \centering
 \includegraphics*[width=15.8cm,viewport=50 82 526 818]{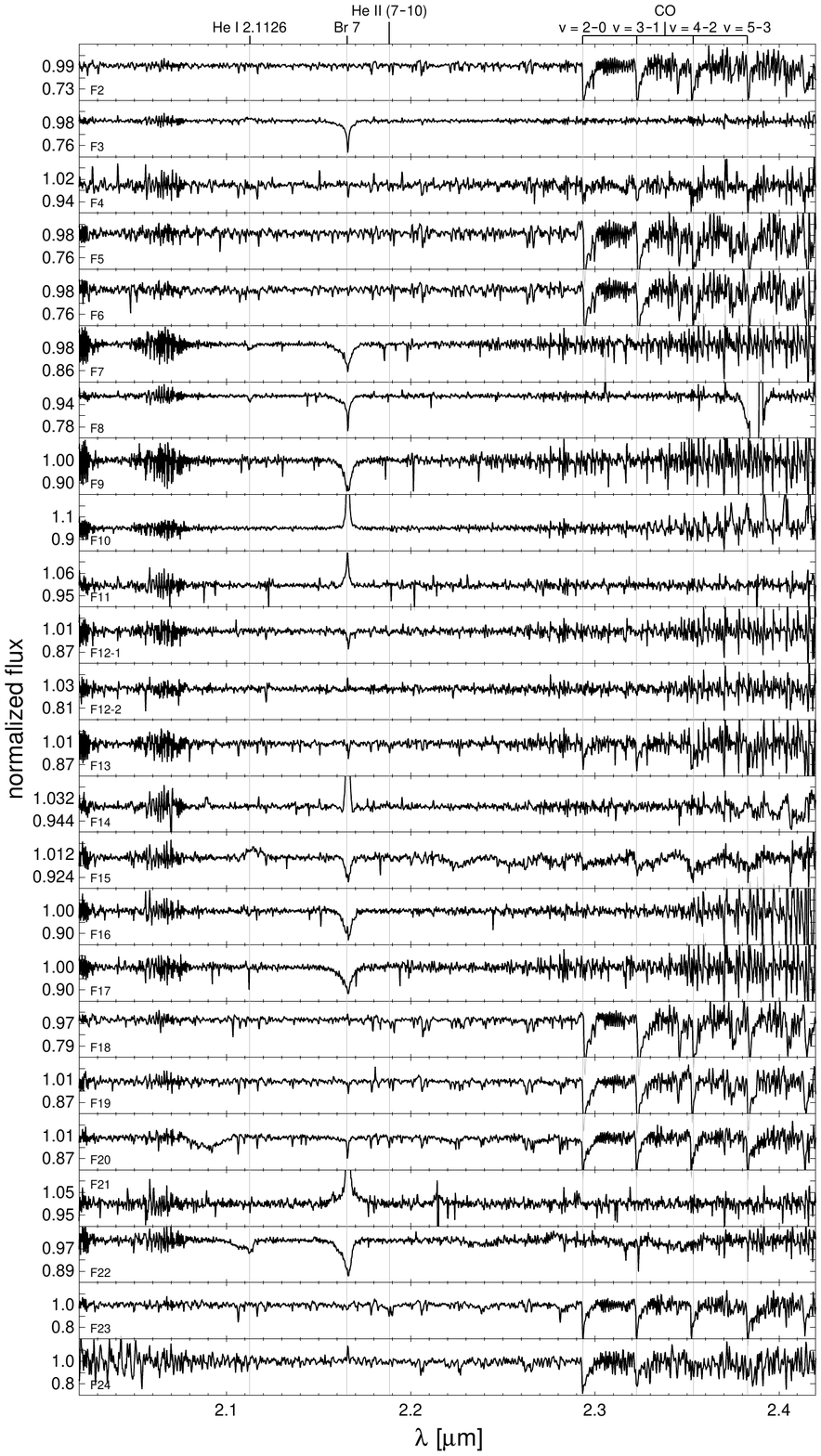}
      \caption{$K$-band spectra in the \textbf{F} sample.}
         \label{fig:spectra-F-K}
\end{figure*}

Figures \ref{fig:spectra-B-H} to ~\ref{fig:spectra-F-K} show the
normalized KMOS spectra of all our targets.

\end{appendix}

\clearpage
\onecolumn

\setcounter{table}{2}
\begin{landscape}
\begin{longtable}{rlrrrrrll}
\caption{Target positions and results.} \label{results.table}\\
\hline
\hline
Star name / position [J2000] & Spectrum & \multicolumn{1}{c}{opt.~mag} & \multicolumn{1}{c}{$J$ [mag]} & \multicolumn{1}{c}{$H$ [mag]} & \multicolumn{1}{c}{$K_s$ [mag]} & \multicolumn{1}{c}{CCCP Match} & \multicolumn{1}{c}{Spectral Type} & \multicolumn{1}{c}{Comment} \\
\hline\noalign{\smallskip}
\endfirsthead
\caption{Continued.} \\
\hline
Star name / position [J2000] & Spectrum & \multicolumn{1}{c}{opt.~mag} & \multicolumn{1}{c}{$J$} & \multicolumn{1}{c}{$H$} & \multicolumn{1}{c}{$K_s$} & \multicolumn{1}{c}{CCCP Match} & \multicolumn{1}{c}{Spectral Type} & \multicolumn{1}{c}{Comment} \\
\hline\noalign{\smallskip}
\endhead
\hline
\endfoot
\hline  
\endlastfoot
2MASS J10444745--5944572  & C16    &$G=16.01$& 10.28 &  8.71 &  7.95 & ...                & cool &  in Tr~16  \\
2MASS J10445220--5944212  & C17    &$G=12.48$& 11.09 & 10.45 &  9.69 & ...                & \textbf{(Ae)Be} &  in Tr~16; CO emission  \\
2MASS J10445360--5947498  & C18    &$G=14.42$& 11.05 & 10.42 & 10.27 & ...            & cool & in Tr~16, foreground star \\
2MASS J10450845--5950054  & F2     &  ...  & 15.73 & 12.30 & 10.72 & ...                & cool &    \\
2MASS J10450879--5950537  & F3     &$G=17.24$& 13.03 & 11.93 & 11.37 & ...            & \textbf{O6--7} &   \\
2MASS J10451232--5949490  & F4     &$G=18.90$& 15.14 & 13.14 & 11.52 & ...                & $\ge$ B2 &   \\
2MASS J10451278--5947511  & F5 + C19  &  ...  & 15.31 & 12.60 & 11.22 & ...                & cool &   \\
2MASS J10451459--5948190  & F6     &  ...  & 15.68 & 12.99 & 11.74 & ...                & cool &   \\
2MASS J10451717--5947013  & F7 + C20 &$G = 16.02$& 12.05 & 11.10 & 10.61 & 104517.21--594701.6 &  \textbf{O9--B1} & OBc~48 \\
2MASS J10451758--5948209  & B3 + C21 &$G=11.82$ & 10.22 &  9.74 &  9.59 & ...              & cool &  foreground star  \\
2MASS J10452013--5950104  & F8     &$G=15.49$& 12.60 & 11.96 & 11.59 & ...            &  \textbf{O8--9}   &  \\
2MASS J10452160--5948457  & B4 + C22 &$G=11.91$& 10.75 & 10.41 & 10.40 & ...              & AFG? &  foreground star \\
2MASS J10452227--5950470  & B5     &$G=10.77$&  8.62 &  8.19 &  7.88 & 104522.29--595047.0 & \textbf{O6} &  MJ~568, OBc~50 \\
2MASS J10452586--5949540  & B6     &$G=19.61$& 13.32 & 11.14 & 10.17 & ...                & cool &   \\
2MASS J10452648--5946188  & F9 + C23 &$G=15.50$& 12.55 & 11.91 & 11.52 & ...          & \textbf{B1--2}   &   \\
2MASS J10452862--5947553  & B7 + C24 &$G=17.50$& 11.91 & 10.28 &  9.46 & 104528.60--594756.1 & \textbf{O6--7} &  OBc~51  \\
2MASS J10452898--5947157  & F10    &$i=20.26$& 15.03 & 13.03 & 11.75 & ...                & \textbf{Be} &   \\
2MASS J10452952--5950390  & F11    &$G=18.88$& 15.01 & 13.18 & 11.76 & ...            & \textbf{(Ae)Be} &  \\
2MASS J10453024--5948206  & B8     &$G=15.32$& 10.43 &  9.24 &  8.60 & 104530.22--594821.0 & \textbf{O5--6} &  OBc~52 \\
VCNS J104530.29--594824.5 & F12-1  &$G=19.67$& 14.21 & 12.78 & 11.94 & ...                & B? &   \\
HAWK-I J104530.42--594825.4 & F12-2  &$i=20.68$& 15.95 & 14.14 & 13.14 & ...                & \textbf{(Ae)Be} &   \\ 
2MASS J10453045--5946467  & B9     &$G=19.89$& 13.13 & 10.71 &  9.68 & ...                & cool &   \\
2MASS J10453053--5948317  & B10    &$G=19.16$& 12.55 & 10.39 &  9.39 & ...                & cool &   \\
2MASS J10453095--5950451  & B11    &$G=16.12$& 12.59 & 11.32 & 10.39 & 104530.96--595045.4 & \textbf{(Ae)Be}   &  \\
2MASS J10453166--5946467  & F13    &$G=19.21$& 14.02 & 12.45 & 11.76 & ...                & cool(?) &  \\
2MASS J10453372--5949596  & F14    &$G=14.20$& 11.78 & 11.34 & 10.97 & 104533.73--595000.5 & \textbf{(Ae)Be} &  \\
2MASS J10453470--5947537  & B12    &$G=17.18$& 11.99 & 10.65 &  9.91 & ...                & \textbf{O9--B1} &   \\
2MASS J10453556--5949314  & B1     &$G=16.08$&  9.81 &  7.84 &  6.90 & 104535.76--594932.7 & cool &   \\
2MASS J10453631--5948233  & B13    &$G=11.71$&  9.29 &  8.81 &  8.51 & 104536.33--594823.5 & \textbf{O5} & MJ~596 = V662~Car \\
2MASS J10453674--5947020  & B14    &$G=13.59$&  9.76 &  8.88 &  8.33 & 104536.75--594702.2 & \textbf{O5}   &  OBc~56 \\
2MASS J10453739--5950310  & B15    &$G=13.06$&  9.49 &  8.26 &  7.85 & ...            & cool &   \\
2MASS J10453745--5949161  & F15    &$G=14.68$& 12.40 & 11.91 & 11.57 & 104537.48--594916.5 & Be? & He~I  emission + CO absorption \\
2MASS J10453755--5948529  & F16    &$G=15.41$& 12.81 & 12.25 & 11.97 & ...            & \textbf{B0--1}  & \\
2MASS J10453774--5947552  & F17    &$G=17.22$& 13.18 & 12.19 & 11.68 & ...            & \textbf{B0--1} &   \\
2MASS J10453806--5947052  & B16    &$G=18.24$& 13.35 & 11.73 & 10.41 & ...            & \textbf{(Ae)Be} &   \\
2MASS J10454029--5948297  & B17    &$G=15.16$& 11.81 & 11.04 & 10.52 & 104540.44--594830.6 & \textbf{(Ae)Be}   &  CO emission\\
2MASS J10454085--5951048  & F18    &$i=20.62$& 14.51 & 12.17 & 11.17 & ...                & cool &  \\
2MASS J10454118--5948587  & F19    &$G=16.04$& 12.68 & 11.42 & 10.78 & ...            & cool &  \\
2MASS J10454173--5951180  & F20    &$G=16.92$& 13.10 & 12.01 & 11.58 & ...            & cool &  \\
2MASS J10454344--5949075  & F21    &$G=19.38$& 15.86 & 14.29 & 12.24 & ...            & \textbf{(Ae)Be} &   \\  
2MASS J10454460--5950411  & B18    &$G=11.85$&  9.70 &  9.30 &  9.06 & ...            & \textbf{O9--B1} &    \\
2MASS J10454559--5946184  & B19    &$G=19.23$& 12.31 &  9.94 &  8.84 & ...            & cool &    \\
2MASS J10454569--5950227  & B20    &$G=16.96$& 11.07 &  9.18 &  8.34 & ...            & cool &   \\
2MASS J10454595--5949075  & F22    &$G=14.82$& 12.21 & 11.69 & 11.42 & ...            & \textbf{O9--B1}   &  \\
2MASS J10454606--5950460  & F23    &$G=13.82$& 11.54 & 10.87 & 10.65 & ...            & cool & foreground star  \\
2MASS J10454661--5948404  & B21    &$G=14.63$& 11.59 & 10.94 & 10.56 & 104546.64--594840.2 & \textbf{O9--B1} & OBc~61 \\
2MASS J10454845--5950343  & B22    &$G=11.25$& 10.18 &  9.89 &  9.84 & ...            & B/A? &  foreground star \\
2MASS J10455242--5949073  & B23    &$G=11.20$& 10.83 & 10.81 & 10.78 & ...            & B? &  foreground star \\
2MASS J10455458--5948150  & B24    &$r=17.58$&  9.30 &  7.76 &  6.74 & ...            & cool & = pillar tip star \\   
HAWK-I J104554.77--594815.5 & F24  &$G=13.35$& 12.38 &  ...  & 11.66 & ...           & cool & MJ~633, foreground star \\
\end{longtable}
\tablefoot{
Optical magnitudes are from Gaia EDR3 ($G$), if available,
 or are $r$- or $i$-band values from the \textit{VST Photometric H-$\alpha$ Survey
of the Southern Galactic Plane} \citep[see][]{Drew14}. 
Near-infrared magnitudes are from our \textit{VISTA Carina Nebula Survey}
\citep{VISTA1} and HAWK-I \citep{HAWKI-survey}.
X-ray source matches refer to the source catalog of the
\textit{CCCP} \citep{CCCP-intro}.
}
\end{landscape}

\end{document}